\documentclass[aps,prd,showpacs,twocolumn]{revtex4}

\usepackage{epsfig}
\usepackage{longtable}

\begin{document}

\begin{flushleft}
ADP-04-17/T599 \\
\end{flushleft}

\title{Spectroscopy of $B_c$ Mesons in the Relativized Quark Model} 
\author{ Stephen Godfrey\footnote{Email: godfrey@physics.carleton.ca}}
\affiliation{
Ottawa-Carleton Institute for Physics, 
Department of Physics, Carleton University, Ottawa, Canada K1S 5B6 
\\ and \\
Special Research Centre for the Subtatomic Structure of Matter\\
University of Adelaide, Adelaide South Australia 5000, Australia }

\date{\today}

\begin{abstract}
We calculate the spectrum of the charm-beauty mesons using the 
relativized quark model.  Using the wavefunctions from this model we 
compute the radiative widths of excited $c\bar{b}$ states. 
The hadronic transition rates between $c\bar{b}$ states are estimated
using the Kuang-Yan approach and are combined with the radiative 
widths to give estimates of the relative branching ratios.
These results are combined with production rates at the Tevatron and 
the LHC to suggest promising signals for excited $B_c$ states.  
Our results are compared with other models to gauge 
the reliability of the predictions and point out differences.

\end{abstract}

\maketitle

\section{Introduction}

The charm-beauty ($B_c$) quarkonium states provide a unique window into heavy quark 
dynamics and therefore an important test of quantum chromodynamics.  
Although they are intermediate to the 
charmonium and bottomonium systems the properties of $B_c$ mesons
are a special case in quarkonium 
spectroscopy as they are the only quarkonia consisting of heavy quarks 
with different flavours. 
Because they carry flavour they 
cannot annihilate into gluons so are more stable 
with widths less than a hundred keV.
Excited $B_c$ states lying below $BD$ (and $BD^*$ or 
$B^*D$) threshold can only undergo radiative or hadronic transitions 
to the ground state $B_c$ 
which then decays weakly.
This results in a rich spectroscopy of  narrow radial and orbital 
excitations below  $B^{(*)}D^{(*)}$ 
threshold 
which are more stable than their charmonium and bottomonium analogues:
There are two sets of $S$-wave states, as many as two $P$-wave 
multiplets (the 1P and some or all of the 2P)
and one $D$-wave multiplet below $BD$ threshold.  As well, the 
$F$-wave multiplet is sufficiently close to threshold that they may 
also be relatively narrow due to angular momentum barrier suppression 
of the Zweig allowed strong decays.

The discovery of the $B_c$ meson by the 
Collider Detector at Fermilab (CDF) Collaboration \cite{abe1998}
in $p\bar{p}$ collisions at $\sqrt{s}=1.8$~TeV has demonstrated the 
possibility of the experimental study of this system and has 
stimulated considerable interest in $B_c$ spectroscopy 
\cite{eichten94,Anikeev:2001rk,gershtein95,gershtein95b,godfrey85,bcspec,zeng95,gupta96,fulcher99,gouz02,ebert03,Kwong:1990am,Ikhdair:2004tj,Brambilla:2000db,Penin:2004xi}.
Calculations of $B_c$ cross sections at hadron colliders
predict that 
large samples of $B_c$ states should be produced at the Tevatron 
and at the LHC opening up this new spectroscopy 
\cite{braaten96,chang03,Chang:1994aw,chang96,cheung93,chang93,Chen93,cheung96,Yuan:1994hn,Cheung:1995ir,Kolodziej:1995nv}. 
At the Tevatron
it is estimated that ${\cal O}(10^7)$ $B_c$ mesons should be produced 
for 1~fb$^{-1}$ of integrated luminosity while 
at the LHC  ${\cal O}(10^9)$  $B_c$ 
mesons are expected to be produced for $L=100$~fb$^{-1}$.  
These numbers are highly sensitive to the $p_T$ and rapidity cuts used to 
extract the signal \cite{chang03,Chang:1994aw}.
The $B_c^*$ cross sections are expected to be 50-100\% larger than the 
$B_c$ cross sections \cite{chang96,cheung93,cheung96}
and Cheung and Yuan \cite{cheung96,Yuan:1994hn} predict 
excited P-waves  to contribute 20\% of inclusive $B_c$ production while
D-wave states are expected to contribute about 2\%
\cite{Cheung:1995ir}.   Chang and Chen 
\cite{chang93}  and Cheung \cite{cheung93} 
estimate that the $2S$ states will be produced in roughly 
the ratio of $2S/1S\simeq 0.6$.  
It should therefore be possible 
to start exploring $c\bar{b}$ spectroscopy at the Tevatron, producing 
the $1P$ and $2S$ states and possibly even some $1D$ and $2P$ states with 
sufficient rate to be observed.  At the LHC, with its higher 
luminosity, the $D$-wave $c\bar{b}$ states should be produce in a 
sizable number so that the LHC should allow the study of the 
spectroscopy and decay of $B_c$ mesons.

In this paper we study the spectroscopy,
including radiative transitions, of charm-beauty mesons using 
the relativized quark model \cite{godfrey85,godfrey85b,godfrey}.  
The model includes one-gluon exchange with a
running coupling constant and a linear confining potential.
It uses relativistic kinematics and momentum dependent
and nonlocal interactions. Although this model is not a 
rigorous calculation from first principles
it gives a good account of most known mesons and 
baryons with only a few free parameters
\cite{godfrey85,godfrey85b,godfrey,capstick,cgip}
so that it provides a useful guide to missing states.
We compare its predictions to those of other 
models with the aim of highlighting which predictions are most 
sensitive to details of the models and therefore the most useful in 
distinguishing models.  However, we are also interested in 
pointing out which predictions give the greatest agreement between 
models and therefore offer the most robust 
signatures for experiments to look for.
Observation of these states and 
measurement of their properties would provide valuable information 
distinguishing details of the various models.

We start with a brief outline of the relativized quark model and 
comment on its similarities and differences with other quark model 
calculations.  The spectroscopic predictions are given and compared to 
those of other models in the literature.  This is followed by
predictions 
for E1 and M1 radiative transitions and estimates of
hadronic transitions based on the Kuang-Yan approach \cite{Yan80,Kua81}.  
We summarize existing predictions for some of the more prominent 
weak decays of the $B_c$ 
ground state as final states are an important 
ingredient in reconstructing the $B_c$ mesons.  In 
addition, leptonic decays measure the wavefunction at the origin and 
are therefore an additional test of the model \cite{Capstick:1989ra}.  
We end by 
discussing some strategies for searching for excited $B_c$ mesons and 
studying their spectroscopy.

\section{Spectroscopy}

In this section we give the 
mass predictions of the relativized quark 
model \cite{godfrey85,godfrey85b,godfrey} 
for the charm-beauty 
mesons and compare those predictions with the predictions of other
calculations. 
This model has ingredients common to many quark potential models
\cite{ebert03,gershtein95,fulcher99,gupta96,zeng95}.  
Almost all such models are based on some variant of the Coulomb plus 
linear potential expected from QCD. An interesting observation is
that all recent models have arrived at the same slope for the
linear confining potential of $\sim 0.18$~GeV$^2$. 
 Most models, as does ours, also include the 
running constant of QCD, $\alpha_s(Q^2)$. And finally, relativistic 
effects are often included at some level.  The relativized quark 
model has been reasonably successful in describing most known mesons. 
Although cracks have recently appeared with the discovery of the 
$D_{sj}$ \cite{Aubert:2003fg,Besson:2003cp,Krokovny:2003zq} 
and $X(3872)$ states \cite{Choi:2003ue}, 
these point to the need to
include physics which has hitherto been neglected such as 
coupled channel effects \cite{Eichten:2004uh}.  

In the relativized quark model 
mesons are approximated by the $q\bar{q}$ sector of
Fock space, in effect integrating out the degrees of freedom below some 
distance scale, $\mu^{-1}$.  
This results in an effective potential, $V(\vec{p},\vec{r})$, 
whose dynamics are governed by a Lorentz vector 
one-gluon-exchange interaction at short distance and
a Lorentz scalar linear confining interaction.  
The basic equation of the model is the 
rest frame Schr\"odinger-type equation \cite{salpeter}:
\begin{equation}
H\vert \psi \rangle = [ H_0 + V_{q\bar{q}} (\vec{p},\vec{r})] \vert \psi
\rangle = E \vert \psi \rangle 
\end{equation}
where
\begin{equation}
H_0 = \sqrt{p^2 +m_q^2} + \sqrt{p^2 + m_{\bar{q}}^2 }
\end{equation}
The effective quark-antiquark potential, $V_{q\bar{q}} (\vec{p},\vec{r})$,
was found by equating the scattering amplitude of free quarks,
using a scattering kernel with the desired Dirac structure, with the effects
between bound quarks inside a hadron \cite{gromes,berestetskij}.
Due to relativistic effects the potential is
momentum dependent in addition to being co-ordinate dependent.
To first order in $(v/c)^2$, $V_{q\bar{q}} (\vec{p},\vec{r})$
reduces to the standard non-relativistic result:
\begin{equation}
V_{q\bar{q}} (\vec{p},\vec{r}) \to V(\vec{r}) =
H^{conf}_{q\bar{q}}  +H^{cont}_{q\bar{q}} + H^{ten}_{q\bar{q}}
+ H^{s.o.}_{q\bar{q}} 
\end{equation}
where
\begin{equation}
H^{conf}_{q\bar{q}} = C + br - {4\over 3} {{\alpha_s(r)} \over r}
\end{equation}
includes the spin-independent linear confinement and Coulomb-like
interaction,
\begin{equation}
H^{cont}_{q\bar{q}} =  {{ 32\pi} \over 9} 
{{ \alpha_s(r)} \over{m_{q}m_{\bar{q}} } } \;
\vec{S}_q \cdot \vec{S}_{\bar{q}} \;
\delta^3 (\vec{r}) \; 
\end{equation}
is the colour contact interaction,
\begin{equation}
H^{ten}_{q\bar{q}} = {4\over 3}{ {\alpha_s(r)}\over {m_q m_{\bar{q}} } } {1\over{r^3}}
\left[{ 
{ { 3\vec{S}_q\cdot \vec{r} \; \vec{S}_{\bar{q}} \cdot \vec{r} } \over {r^2} } 
- \vec{S}_q \cdot \vec{S}_{\bar{q}} 
}\right] \; \; 
\end{equation}
is the colour tensor interaction,
\begin{equation}
H^{s.o.}_{q\bar{q}} = H^{s.o.(cm)}_{q\bar{q}} + H^{s.o.(tp)}_{q\bar{q}}
\end{equation}
is the spin-orbit interaction with
\begin{equation}
\label{eqn:socm}
H^{s.o.(cm)}_{q\bar{q}} ={{4\alpha_s(r)}\over { 3 r^3}}
\left({ 
{ {\vec{S}_q} \over {m_q m_{\bar{q}} }} 
+ { {\vec{S}_{\bar{q}} }\over {m_q m_{\bar{q}}} } 
+ { {\vec{S}_q} \over {m_q^2 } } 
+ { {\vec{S}_{\bar{q}} }\over {m_{\bar{q}}^2 } } 
}\right) 
\cdot \vec {L} 
\end{equation}
its colour magnetic piece arising from one-gluon exchange and
\begin{equation}
\label{eqn:sotp}
H^{s.o.(tp)}_{q\bar{q}} =- {1\over{2r}}
{ { \partial H_{q\bar{q}}^{conf} } \over{\partial r} }
\left({ { {\vec{S}_q} \over {m_q^2 } } 
+ { {\vec{S}_{\bar{q}} }\over {m_{\bar{q}}^2 } } }\right) \cdot \vec {L}
\end{equation}
the Thomas precession term.  In these formulae
$\alpha_s(r)$ is the running coupling constant of QCD.

To relativize the $q\bar{q}$
potential, the full Dirac scattering amplitude was used as a starting
point which for on-shell $q\bar{q}$ scattering is exact.  However for 
a strongly interacting system there will in general be off-shell 
behavior which we did not consider in addition to other simplifications
such as neglecting more complex components of Fock space.  We therefore
built a semiquantitative model of relativistic effects by smearing
the co-ordinate $\vec{r}$ 
over the distances of the order of the inverse quark mass
by convoluting the potential with a Gaussian form factor and 
replacing factors
of $m_i^{-1}$  with, roughly speaking, factors of $(p^2+m_i^2)^{-1/2}$. 
 The details of this {\it relativization}
procedure and the method of solution can be found in Ref. 
\cite{godfrey85}.  It should be kept in mind 
that because we neglected coupled channel 
effects and the crudeness of the relativization procedure we do not 
expect the mass predictions to be accurate to better than $\sim 
10-20$~MeV.

For the case of a quark and antiquark of unequal mass charge conjugation
parity is no longer a good quantum number so that states with different 
total spins but with the same total angular momentum, such as
the $^3P_1 -^1P_1$ and $^3D_2 -^1D_2$ pairs, can mix via
the spin orbit interaction or some other mechanism.
Eqns. \ref{eqn:socm} and \ref{eqn:sotp} 
can be rewritten to explicitely give the antisymmetric 
spin-orbit mixing term:
\begin{equation}
\label{eqn:somix}
H_{s.o.}^- =
+ {1\over 4} \left({ {4\over 3} {{\alpha_s}\over{r^3}} - {b \over r} } \right)
 \left( { {1\over{m_Q^2}} - {1\over {m_{\bar{Q}}^2}} }\right) 
\; \vec{S}_-\cdot \vec{L}
\end{equation}
where $\vec{S}_- = \vec{S}_Q -\vec{S}_{\bar{Q}}$.
Consequently, the physical $j=1$ $P$-wave states are linear
combinations of $^3P_1$ and $^1P_1$ which we describe by:
\begin{eqnarray}
\label{eqn:mixing}
P'  & = ^1P_1 \cos\theta_{nP} + ^3P_1 \sin\theta_{nP} \nonumber \\   
P & =-^1P_1 \sin\theta_{nP} + ^3P_1 \cos \theta_{nP} 
\end{eqnarray}
with analogous notation for the corresponding $L=D$, $F$, etc. pairs.
In eqn. \ref{eqn:mixing}
$P\equiv L=1$ designates the relative angular momentum of the $Q\bar{Q}$ 
pair and the subscript $J=1$ is the total angular momentum of the $Q\bar{Q}$ 
pair which is equal to $L$.  Our notation implicitely implies $L-S$ 
coupling between the quark spins and the relative orbital angular momentum.  
In the heavy quark limit in which 
the heavy quark mass $m_Q\to \infty$, 
the states can be described by the total angular momentum of the
light quark, $j$, which couples to the spin of the heavy quark and
corresponds to $j-j$ coupling.  This limit gives 
rise to two doublets, one with $j=1/2$ and the 
other with 
$j=3/2$ and corresponds to two physically independent mixing angles 
$\theta=-\tan^{-1}(\sqrt{2})\simeq -54.7^\circ$ and 
$\theta=\tan^{-1}(1/\sqrt{2})\simeq 35.3^\circ$ \cite{barnes}. Some 
authors prefer to use the $j-j$ basis \cite{eichten94}
but since we solve our 
Hamiltonian equations assuming $L-S$ eigenstates and then include the 
$LS$ mixing we use the notation of eqn. \ref{eqn:mixing}.  
It is straightforward to
transform between the $L-S$ basis and the $j-j$ basis.  It will turn 
out that radiative transitions are particularly sensitive to the 
$^3L_L-^1L_L$ mixing angle with predictions from different 
models in some cases
giving radically different results.  We also note that the 
definition of the mixing angles are frought with ambiguities.  For 
example, charge conjugating $c\bar{b}$ into $b\bar{c}$ flips the 
sign of the angle and the phase convention depends on the
order of coupling $\vec{L}$, $\vec{S}_Q$ and $\vec{S}_{\bar{Q}}$
\cite{barnes}.

The  Hamiltonian problem was solved 
using the following parameters: the slope of the
linear confining potential is 0.18 GeV$^2$, 
$m_c=1.628$ GeV, and $m_b=4.977$ GeV.  The predictions of our model 
are given in Fig. 1 and are compared to the predictions of other 
calculations in Table \ref{tab:masses}.  
Because the mixing angles defined in eqn. \ref{eqn:mixing} are 
important for predictions of radiative transitions those predictions
are also given in Table \ref{tab:masses}.  Although  I have 
attempted to consistently give the masses and the mixing angles 
of the predicted eigenstates in the convention of eqn. 
\ref{eqn:mixing}, because not all 
authors have unambiguously  defined their phase conventions  I 
cannot guaranty that these results are free of inconsistantcies.

The different models are in remarkable agreement with the differences, 
for the most part, within the expected accuracy of the models.  This 
almost certainly indicates how the various models have converged 
to using similar confining potentials and including a strong 
running coupling constant in the Coulomb piece of the potential.  The 
only significant difference is the larger spread ($\sim 70$~MeV) 
for the $1D$ multiplet centre of gravity predictions. 
The spin-dependent splittings 
are also in reasonable agreement.  Potential models can therefore be 
used as a reliable guide in searching for the  $B_c$ excited states.
An important difference in the predictions 
is that in the Eichten-Quigg calculation \cite{eichten94} 
the $1P_1$ 
states are almost pure $^3P_1$ and $^1P_1$ with little mixing while in 
other models there is significant mixing.  
This arises from the much smaller expectation value of the off-diagonal 
mixing term (eqn. \ref{eqn:somix}) in the Eichten-Quigg calculation 
\cite{eichten94}  compared to the other models.  
Since, after rotating from 
the $j-j$ basis to the $L-S$ basis, the $L-S$ mixing term given by 
Eichten and Quigg is in agreement with eqn. \ref{eqn:somix}, the 
differences in expectation values can only be 
attributed to differences in the cancellations between the short 
and long distance pieces in eqn. \ref{eqn:somix},  i.e. between the 
$\frac{4}{3}\frac{\alpha_s}{r^3}$ and $\frac{b}{r}$ pieces.  This 
reflects subtle differences in the $q\bar{q}$ potentials of different 
models.
Because the E1 radiative transitions are sensitive to the $^3P_1-^1P_1$ 
mixing,
the measurement of radiative transitions can be used to 
distinguish between the different models.  
The study of the $B_c$ spectroscopy will help test and refine the 
quark potential models but more importantly will test Lattice QCD, 
NRQCD, and pNRQCD etc. which are more directly connected to QCD.

\begin{table*}
\caption{Predicted masses and Spin-Orbit mixing angles.  The first 
column labelled GI is the present work.
The $P_1' -P_1$, $D_2' -D_2$, and $F_3' -F_3$ states and mixing angles 
are defined using the convention of eqn. \ref{eqn:mixing}. 
\label{tab:masses}}
\begin{tabular}{lcccccccc} \hline
State 
	& GI\cite{godfrey85} 
	& EFG \cite{ebert03} 
	& FUII \cite{fulcher99}
        & GKLT \cite{gershtein95b}
        & EQ \cite{eichten94} 
	& GJ \cite{gupta96} 
	& ZVR \cite{zeng95} 
	& Lattice \cite{davies96}\footnotemark[1] \\ 
\hline
$1^3S_1 $ & 6338 & 6332	& 6341 & 6317 & 6337 & 6308 & 6340 & $6321\pm20$\\
$1^1S_0 $ & 6271 & 6270 & 6286 & 6253 & 6264 & 6247 & 6260 & 
						 $6280\pm30 \pm 190$\\ \hline
$1^3P_2 $ & 6768 & 6762	& 6772 & 6743 & 6747 & 6773 & 6760 & $6783\pm 30$ \\
$1 P_1' $ & 6750 & 6749 & 6760 & 6729 & 6736 & 6757 & 6740 & $6765\pm 30$\\
$1 P_1 $  & 6741 & 6734 & 6737 & 6717 & 6730 & 6738 & 6730 & $6743\pm 30$\\
$1^3P_0 $ & 6706 & 6699 & 6701 & 6683 & 6700 & 6689 & 6680 & $6727\pm 30$\\
$\theta_{1P}$ & 22.4$^\circ$ & 20.4$^\circ$ & 28.5$^\circ$  
		& 17.1$^\circ$  & $\sim -2^\circ$ & 25.6$^\circ$ 
					& & $33.4\pm 1.5^\circ$ \\ \hline
$2^3S_1 $ & 6887 & 6881	& 6914 & 6902 & 6899 & 6886 & 6900 & $6990\pm 80$ \\
$2^1S_0 $ & 6855 & 6835 & 6882 & 6867 & 6856 & 6853 & 6850 & $6960\pm 80$\\ \hline
$2^3P_2 $ & 7164 & 7156 &	& 7134 & 7153 &	     & 7160 & \\
$2 P_1' $ & 7150 & 7145 &	& 7124 & 7142 &      & 7150 & \\
$2 P_1 $  & 7145 & 7126 &	& 7113 & 7135 &      & 7140 & \\
$2^3P_0 $ & 7122 & 7091 &	& 7088 & 7108 &	     & 7100 & \\ 
$\theta_{2P}$ & 18.9$^\circ$ & 23.2$^\circ$ & 	
		& 21.8$^\circ$  & $-17^\circ$  & 	&   & \\ \hline
$3^3S_1 $ & 7272 & 7235 &	& 	& 7280 &     & 7280 & \\
$3^1S_0 $ & 7250 & 7193 &	&	& 7244 &     & 7240 & \\ \hline
$1^3D_3 $ & 7045 & 7081 & 7032	& 7007	& 7005 &     & 7040 & \\
$1 D_2' $ & 7036 & 7079 & 7028	& 7016	& 7009\footnotemark[2]  &     & 7030 & \\
$1 D_2 $  & 7041 & 7077 & 7028 	& 7001	& 7012\footnotemark[2] &     & 7020 & \\
$1^3D_1 $ & 7028 & 7072 & 7019  & 7008	& 7012 &     & 7010 & \\
$\theta_{1D}$ & 44.5$^\circ$ & -35.9$^\circ$ 	& 	& -34.5$^\circ$
				        & 	&    &      & \\   \hline
$1^3F_4 $ & 7271 &	&	&	& 	&    & 7250 & \\
$1 F_3' $ & 7266 &	&	&	&	&    & 7250 & \\
$1 F_3 $  & 7276 &	&	&	&	&    & 7240 & \\
$1^3F_2 $ & 7269 &	&	&	&	&    & 7240 & \\
$\theta_{1F}$ & 41.4$^\circ$ & 	& 	&  &  &   & & \\
\hline
\end{tabular}
\footnotetext[1]{The error estimates are taken from Ref. 
\cite{fulcher99}.}
\footnotetext[2]{We identify the $1D_2'$ and $1D_2$ states with the 
$^1D_2$ and $^3D_2$ states of Eichten and Quigg \cite{eichten94}. }
\end{table*}

\begin{figure*}[t]
\begin{center}
\centerline{\epsfig{file=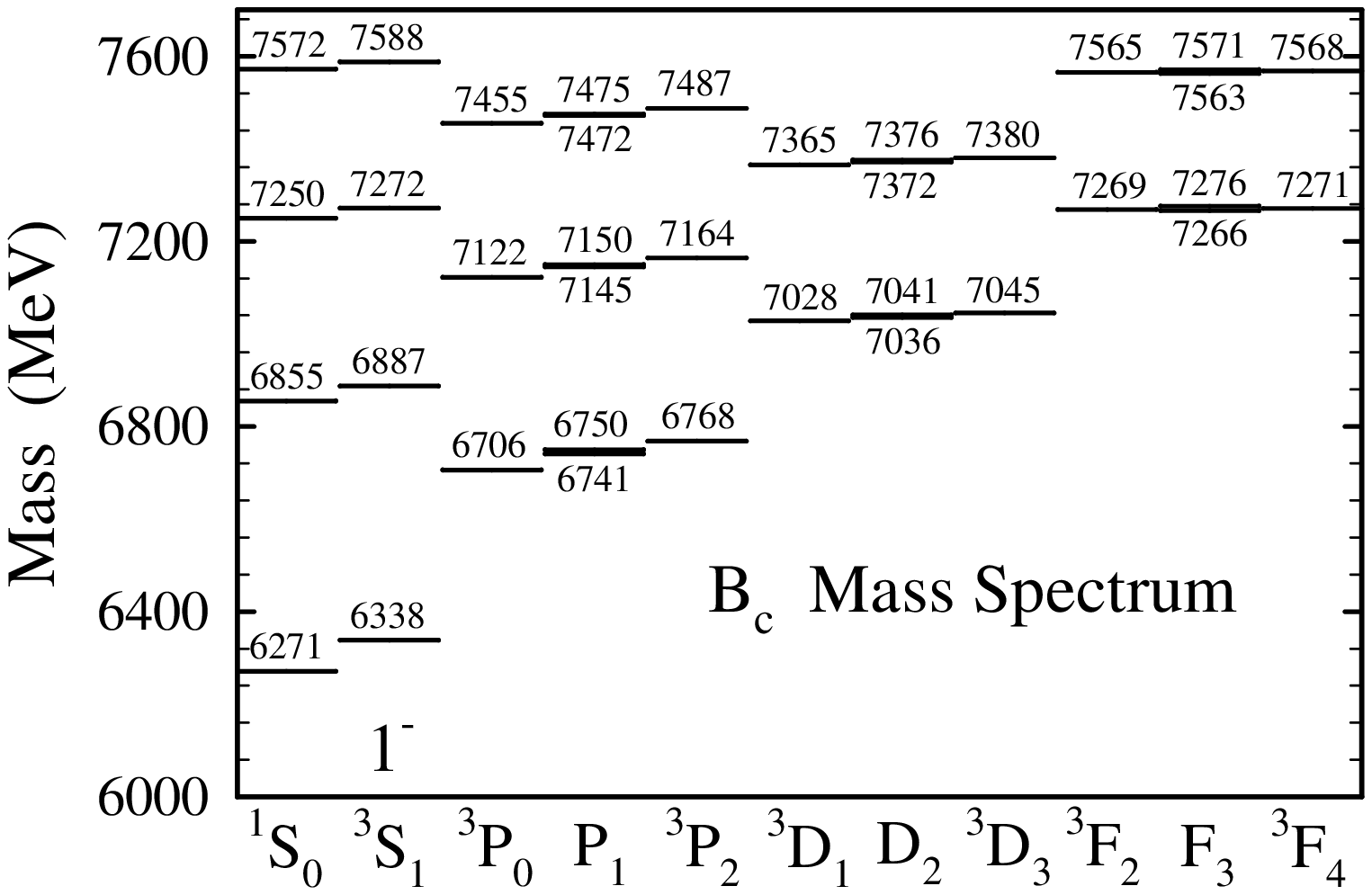,width=5.5in,clip=}}
\end{center}
\caption{The $B_c$ mass spectrum. }
\label{Fig1}
\end{figure*}

\section{Radiative Transitions}

Radiative transitions will likely play an important role in the discovery 
and identification of $B_c$ states.  
In this section we calculate the E1 and M1 radiative widths.
The partial width for an E1 radiative transition between
states in the nonrelativistic quark model is given by 
\cite{Kwo88}
\begin{widetext}
\begin{equation}
\Gamma( 
{\rm n}\, {}^{2{S}+1}{\rm L}_{J} 
\to 
{\rm n}'\, {}^{2{S}'+1}{\rm L}'_{{J}'}  
+ \gamma) 
 =  \frac{4}{3}\,  \langle e_Q \rangle^2 \, \alpha \,
\omega^3 \,   
C_{fi}\,
\delta_{{S}{S}'} \, 
|\,\langle 
{\rm n}'\, {}^{2{S}'+1}{\rm L}'_{{J}'} 
|
\; r \; 
|\, 
{\rm n}\, {}^{2{S}+1}{\rm L}_{J}  
\rangle\, |^2  
\ ,
\end{equation}
\end{widetext}
where 
\begin{equation}
\langle e_Q \rangle = {{m_b e_c - m_c e_{\bar{b}} }\over{m_b +m_c}}
\end{equation}
$e_c= 2/3$ is the $c$-quark charge and
$e_b= -1/3$ is the $b$-quark charge in units of $|e|$,
$\alpha$ is the fine-structure constant,
$\omega$ is the photon's energy, and $C_{fi}$ is given by
\begin{equation}
C_{fi}=\hbox{max}({L},\; {L}') (2{J}' + 1)
\left\{ { {{L}' \atop {J}} {{J}' \atop {L}} {{S} \atop 1}  } \right\}^2 .
\end{equation}
For convenience the $C_{fi} $ coefficients are listed 
in Table \ref{tab:e1}.  The matrix elements 
$\langle {n'}^{2{S}'+1}{L}'_{{J}'} |\; r \; 
| n^{2{S}+1}{L}_{J}  \rangle$ 
are given in Table \ref{tab:e1} and were evaluated 
using the wavefunctions given by the relativized quark model \cite{godfrey85}.
Relativistic corrections are implicitly included in these E1 
transitions through Siegert's theorem \cite{Sie37,McC83,Mox83}, 
by including spin-dependent interactions in the Hamiltonian used to 
calculate the meson masses and wavefunctions.   
The E1 radiative widths are given in Table \ref{tab:e1} and compared 
to other predictions in Table \ref{tab:e1comp}.

Most of the predictions for E1 transitions are in qualitative
agreement.  While most differences are due to differences in
phase space arising from different mass predictions the 
more interesting differences arise from wavefunction effects. 
The largest differences are for decays involving the $P_1$ 
and $P_1'$ states which are mixtures of the spin singlet $^1P_1$ and 
spin triplet $^3P_1$ states.  These can be traced back to 
the different $^3P_1-^1P_1$mixing angles predicted by the different models.
Wavefunction effects also appear in decays from radially excited 
states to ground state mesons such as $2^3P_0\to 1^3S_1 \; \gamma$ 
which varies from 1 to 22~keV.  Because the $2^3P_0$ has a node in its
wavefunction there will be a cancellation between different pieces of the 
$\langle2^3P_0 | r |^3S_1 \rangle$ overlap integral.
Ebert, Faustov, and Galkin \cite{ebert03}
include an additional relativistic correction to transitions 
involving the mixed states caused by the difference of the $c$ and $b$ 
quark masses.   This leads to further differences with the other 
models.  

\begin{table*}
\caption{E1 transition rates. The matrix elements were obtained using
the wavefunctions of the GI model \cite{godfrey85}. 
For mixed states such as 
$1P_1'$ and $1P_1$ the widths are calculated
using the wavefunction conventions defined in eqn. 
\ref{eqn:mixing} with the mixing angles given in Table I and the 
matrix elements and $C_{fi}$ factors corresponding to the labelling of 
column 6.
\label{tab:e1}}
\begin{ruledtabular}
\begin{tabular}{l l r c c l c c c  } 
Initial & Final & $M_i$ & $M_f$ &  $\omega$ & $i\to f$ &
	$\langle f| r | i \rangle $ & $C_{fi}$ & Width   \\
state & state & (MeV) & (MeV) & (MeV) & & (GeV$^{-1}$) &  & (keV) \\
\hline 
$1^3P_2 $ & $1^3S_1 \; \gamma$ 
	&  6768 & 6338 & 416  & $1^3P_2\to 1^3S_1$ & 1.421 & $\frac{1}{3}$ & 83 \\
$1P_1' $ & $1^3S_1 \; \gamma$ 
	&  6750 & 6338 & 399  & $1^3P_1\to 1^3S_1$ &1.435  & $\frac{1}{3}$ & 11 \\
	 & $1^1S_0 \; \gamma$ 
	&  & 6271 & 462  & $1^1P_1\to 1^1S_0$ & 1.288 & $\frac{1}{3}$ & 80 \\
$1P_1 $ & $1^3S_1 \; \gamma$ 
	&  6741 & 6338 & 391  & $1^3P_2\to 1^3S_1$ & 1.435 & $\frac{1}{3}$ & 60 \\
	 & $1^1S_0 \; \gamma$ 
	&  & 6271 & 454  & $1^1P_1\to 1^1S_0$ & 1.288 & $\frac{1}{3}$ & 13 \\
$1^3P_0 $ & $1^3S_1 \; \gamma$ 
	&  6706 & 6338 & 358  & $1^3P_0\to 1^3S_1$ & 1.443 & $\frac{1}{3}$ & 55 \\ 
\hline
$2^3S_1 $ & $1^3P_2 \; \gamma$ 
	&  6887 & 6768 & 118  & $2^3S_1\to 1^3P_2$ &-1.914 & $\frac{5}{9}$ & 5.7 \\
	& $1P_1' \; \gamma$ 
	&  	& 6750 & 136  & $2^3S_1\to 1^3P_1$ & -1.777 & $\frac{1}{3}$ & 0.7 \\
	& $1P_1 \; \gamma$ 
	&  	& 6741 & 144  & $2^3S_1\to 1^3P_1$ & -1.777 & $\frac{1}{3}$ & 4.7 \\
	& $1^3P_0 \; \gamma$ 
	&  	& 6706 & 179  & $2^3S_1\to 1^3P_0$ & -1.620 & $\frac{1}{9}$ & 2.9 \\
$2^1S_0$ & $1P_1' \; \gamma$ 
	& 6855 	& 6750 & 104  & $2^1S_0\to 1^1P_1$ & -1.909 & 1	 & 6.1 \\
	& $1P_1 \; \gamma$ 
	&  	& 6741 & 113  & $2^1S_0\to 1^1P_1$ & -1.909 & 1	 & 1.3 \\
\hline
$1^3D_3 $ & $1^3P_2 \; \gamma$ 
	&  7045 & 6768 & 272  & $1^3D_3\to 1^3P_2$ & 2.383 & $\frac{2}{5}$ & 78 \\
$1D_2'$ & $1^3P_2 \; \gamma$ 
	&7036   & 6768 & 263   & $1^3D_2\to 1^3P_2$ & 2.389 & $\frac{1}{10}$ & 8.8 \\
	& $1P_1' \; \gamma$ 
	& 	& 6750 & 280  & $1^1D_2\to 1^1P_1$ & 2.306 & $\frac{2}{5}$ & 63 \\
	& $1P_1 \; \gamma$ 
	& 	& 6741 & 289  & $1^1D_2\to 1^1P_1$ & 2.306 & $\frac{2}{5}$ & 7.0 \\
$1D_2 $ & $1^3P_2 \; \gamma$ 
	&  7041 & 6768 & 268  & $1^3D_2\to 1^3P_2$ & 2.389 & $\frac{1}{10}$ & 9.6 \\
	& $1P_1' \; \gamma$ 
	& 	& 6750 & 285  & $1^3D_2\to 1^3P_1$ & 2.274 & $\frac{3}{10}$ & 15 \\
	& $1P_1 \; \gamma$ 
	& 	& 6741 & 294  & $1^3D_2\to 1^3P_1$ & 2.274 & $\frac{3}{10}$ & 64 \\
$1^3D_1 $ & $1^3P_2 \; \gamma$ 
	&  7028 & 6768 & 255  & $1^3D_1\to 1^3P_2$ & 2.391 & $\frac{1}{90}$ & 1.8 \\
	& $1P_1' \; \gamma$ 
	&  	& 6750 & 273  & $1^3D_1\to 1^3P_1$ & 2.281 & $\frac{1}{6}$ & 4.4 \\
	& $1P_1 \; \gamma$ 
	&  	& 6741 & 281  & $1^3D_1\to 1^3P_1$ & 2.281 & $\frac{1}{6}$ & 28 \\
	& $1^3P_0 \; \gamma$ 
	&  	& 6706 & 315  & $1^3D_1\to 1^3P_0$ & 2.152 & $\frac{2}{9}$ & 55 \\
\hline
$2^3P_2 $ & $2^3S_1 \; \gamma$ 
	&  7164 & 6887 & 272  & $2^3P_2\to 2^3S_1$ & 2.195 & $\frac{1}{3}$ & 55 \\
	& $1^3S_1 \; \gamma$ 
	&  	& 6338 & 778  & $2^3P_2\to 1^3S_1$ & 0.2308 & $\frac{1}{3}$ & 14 \\
	& $1^3D_3 \; \gamma$ 
	&  	& 7045 & 118  & $2^3P_2\to 1^3D_3$ & -2.072 & $\frac{14}{25}$ & 6.8 \\
	& $1D_2' \; \gamma$ 
	&  	& 7036 & 127  & $2^3P_2\to 1^3D_2$ & -1.970 & $\frac{1}{10}$ & 0.7 \\
	& $1D_2 \; \gamma$ 
	&  	& 7041 & 122  & $2^3P_2\to 1^3D_2$ & -1.970 & $\frac{1}{10}$ & 0.6 \\
	& $1^3D_1 \; \gamma$ 
	&  	& 7028 & 135  & $2^3P_2\to 1^3D_1$ & -1.866 & $\frac{1}{150}$ & 0.1 \\
$2P_1' $ & $2^3S_1 \; \gamma$ 
	&  7150 & 6887 & 258  & $2^3P_1\to 2^3S_1$ & 2.319 & $\frac{1}{3}$ & 5.5 \\
	& $2^1S_0 \; \gamma$ 
	&  	& 6855 & 289  & $2^1P_1\to 2^1S_0$ & 2.046 & $\frac{1}{3}$ & 52 \\
	& $1^3S_1 \; \gamma$ 
	&  	& 6338 & 769  & $2^3P_1\to 1^3S_1$ & 0.155 & $\frac{1}{3}$ & 0.6 \\
	& $1^1S_0 \; \gamma$ 
	&  	& 6271 & 825  & $2^1P_1\to 1^1S_0$ & 0.254 & $\frac{1}{3}$ & 19 \\
	& $1D_2' \; \gamma$ 
	&  	& 7036 & 113  & $2^3P_1\to 1^3D_2$ & -2.096 & $\frac{1}{2}$ & 5.5 \\
	& $1D_2 \; \gamma$ 
	&  	& 7041 & 108  & $2^1P_1\to 1^1D_2$ & -2.080 & $\frac{2}{3}$ & 1.3 \\
	& $1^3D_1 \; \gamma$ 
	&  	& 7028 & 121  & $2^3P_1\to 1^3D_1$ & -1.996 & $\frac{1}{6}$ & 0.2 \\
$2P_1 $ & $2^3S_1 \; \gamma$ 
	&  7145 & 6887 & 253  & $2^3P_1\to 2^3S_1$ & 2.319 & $\frac{1}{3}$ & 45 \\
	& $2^1S_0 \; \gamma$ 
	&  	& 6855 & 284  & $2^1P_1\to 2^1S_0$ & 2.046 & $\frac{1}{3}$ & 5.7 \\
	& $1^3S_1 \; \gamma$ 
	&  	& 6338 & 761  & $2^3P_1\to 1^3S_1$ & 0.155 & $\frac{1}{3}$ & 5.4 \\
	& $1^1S_0 \; \gamma$ 
	&  	& 6271 & 820  & $2^1P_1\to 1^1S_0$ & 0.254 & $\frac{1}{3}$ & 2.1 \\
	& $1D_2' \; \gamma$ 
	&  	& 7036 & 108  & $2^3P_1\to 1^3D_2$ & -2.096 & $\frac{1}{2}$ & 0.8 \\
	& $1D_2 \; \gamma$ 
	&  	& 7041 & 103  & $2^1P_1\to 1^1D_2$ & -2.080 & $\frac{2}{3}$ & 3.6 \\
	& $1^3D_1 \; \gamma$ 
	&  	& 7028 & 116  & $2^3P_1\to 1^3D_1$ & -1.996 & $\frac{1}{6}$ & 1.6 \\
$2^3P_0 $ & $2^3S_1 \; \gamma$ 
	&  7122 & 6887 & 231  & $2^3P_0\to 2^3S_1$ & 2.437 & $\frac{1}{3}$ & 42 \\
	& $1^3S_1 \; \gamma$ 
	&  	& 6338 & 741  & $2^3P_0\to 1^3S_1$ & 0.066 & $\frac{1}{3}$ & 1.0 \\
	& $1^3D_1 \; \gamma$ 
	&  	& 7028 & 93  & $2^3P_0\to 1^3D_1$ & -2.128 & $\frac{2}{3}$ & 4.2 \\
\end{tabular}
\end{ruledtabular}
\end{table*}

\begin{table*}
\caption{E1 transition rates (continued).
\label{tab:e1b}}
\begin{ruledtabular}
\begin{tabular}{l l r c c l c c c  } 
Initial & Final & $M_i$ & $M_f$ &  $\omega$ & $i\to f$ &
	$\langle f| r | i \rangle $ & $C_{fi}$ & Width   \\
state & state & (MeV) & (MeV) & (MeV) & & (GeV$^{-1}$) &  & (keV) \\
\hline 
$1^3F_4 $ & $1^3D_3 \; \gamma$ 
	&  7271 & 7045 & 222  & $1^3F_4\to 1^3D_3$ & 3.156 & $\frac{3}{7}$ & 81 \\
$1F_3' $ & $1^3D_3 \; \gamma$ 
	&  7266 & 7045 & 218  & $1^3F_3\to 1^3D_3$ & 3.159 & $\frac{1}{21}$ & 3.7 \\
	& $1D_2' \; \gamma$ 
	&  	& 7036 & 226  & $1^1F_3\to 1^1D_2$ & 3.104 & $\frac{3}{7}$ & 78 \\
	& $1D_2 \; \gamma$ 
	&  	& 7041 & 222  & $1^3F_3\to 1^3D_2$ & 3.091 & $\frac{8}{21}$ & 0.5 \\
$1F_3 $ & $1^3D_3 \; \gamma$ 
	&  7276 & 7045 & 227  & $1^3F_3\to 1^3D_3$ & 3.159 & $\frac{1}{21}$ & 5.4 \\
	& $1D_2' \; \gamma$ 
	&  	& 7036 & 236  & $1^1F_3\to 1^1D_2$ & 3.104 & $\frac{3}{7}$ & 0.04 \\
	& $1D_2 \; \gamma$ 
	&  	& 7041 & 231  & $1^3F_3\to 1^3D_2$ & 3.091 & $\frac{8}{21}$ & 82 \\
$1^3F_2 $ & $1^3D_3 \; \gamma$ 
	&  7269 & 7045 & 221  & $1^3F_2\to 1^3D_3$ & 3.160 & $\frac{1}{525}$ & 0.4 \\
	& $1D_2' \; \gamma$ 
	&  	& 7041 & 224  & $1^3F_2\to 1^3D_2$ & 3.095 & $\frac{1}{15}$ & 6.3 \\
	& $1D_2 \; \gamma$ 
	&  	& 7036 & 229  & $1^3F_2\to 1^3D_2$ & 3.095 & $\frac{1}{15}$ & 6.5 \\
	& $1^3D_1 \; \gamma$ 
	&  	& 7028 & 237  & $1^3F_2\to 1^3D_1$ & 3.026 & $\frac{9}{25}$	 & 75 \\
\end{tabular}
\end{ruledtabular}
\end{table*}

\begin{table*}
\caption{Comparison of predictions for E1 transition rates. The column 
labelled GI is the present work. We quote 
the predicted rates from the various references and do not attempt to 
normalize the rates to common phase space factors.
\label{tab:e1comp}}
\begin{ruledtabular}
\begin{tabular}{l l c c c c c c  } 
Initial & Final &  \multicolumn{6}{c}{Widths (keV)} \\
 state & state  & GI \cite{godfrey85}  
		& EFG \cite{ebert03}	 
		& GKLT \cite{gershtein95b}  
		& EQ \cite{eichten94}
		& GJ \cite{gupta96}   
		& FU \cite{fulcher99} \\
\hline 
$1^3P_2 $ & $1^3S_1 \; \gamma$ 
	& 83 & 107  & 102.9 & 112.6 & 73.6 & 126 \\
$1P_1' $ & $1^3S_1 \; \gamma$ 
	&  11 & 13.6 & 8.1 & 0.1 & 10.5 & 26.2 \\
	 & $1^1S_0 \; \gamma$ 
	&  80 & 132 & 131.1 & 56.4 & 66.6 & 128 \\
$1P_1 $ & $1^3S_1 \; \gamma$ 
	&   60 & 78.9 & 77.8 & 99.5 & 49.0 & 75.8  \\
	 & $1^1S_0 \; \gamma$ 
	&  13 & 18.4  & 11.6 & 0.0 & 16.6 &  32.5 \\
$1^3P_0 $ & $1^3S_1 \; \gamma$ 
	&  55  & 67.2  & 65.3 & 79.2 & 43.0 & 74.2 \\ 
\hline
$2^3S_1 $ & $1^3P_2 \; \gamma$ 
	 & 5.7 & 5.18 & 14.8 & 17.7 & 4.0 & 14.5 \\
	& $1P_1' \; \gamma$ 
	 & 0.7 & 0.63 & 1.0 & 0.0 & 0.6 & 2.5 \\
	& $1P_1 \; \gamma$ 
	 & 4.7 & 5.05 & 12.8 & 14.5 & 3.6 & 13.3 \\
	& $1^3P_0 \; \gamma$ 
	 & 2.9 & 3.78 & 7.7 & 7.8 & 2.6 & 9.6 \\
$2^1S_0$ & $1P_1' \; \gamma$ 
	 & 6.1 & 3.72 & 15.9 & 5.2 & 3.6 & 13.1 \\
	& $1P_1 \; \gamma$ 
	 & 1.3 & 1.02 & 1.9 & 0.0 & 1.3 & 6.4 \\
\hline
$1^3D_3 $ & $1^3P_2 \; \gamma$ 
	&  78 & 102 & 76.9 & 98.7 & &\\
$1D_2' $ & $1^3P_2 \; \gamma$ 
	&   8.8 & 12.8  & 6.8 &  & &\\
	& $1P_1' \; \gamma$ 
	&  63 & 116  & 46.0 & 92.5 \footnotemark[1] & & \\
	& $1P_1 \; \gamma$ 
	&  7.0 & 7.25 & 25.0 &  & &\\
$1D_2 $	& $1^3P_2 \; \gamma$ 
	& 9.6  & 27.5  & 12.2 & 24.7 \footnotemark[1] & & \\
	& $1P_1' \; \gamma$ 
	& 15 & 14.1 & 18.4 & 0.1 \footnotemark[1] & &\\
	& $1P_1 \; \gamma$ 
	&  64 & 112 & 44.6 & 88.8 \footnotemark[1] & & \\
$1^3D_1 $ & $1^3P_2 \; \gamma$ 
	&  1.8 & 5.52 & 2.2 & 2.7 & & \\
	& $1P_1' \; \gamma$ 
	& 4.4 & 7.66 & 3.3 & 0.0 & &\\
	& $1P_1 \; \gamma$ 
	& 28 & 73.8 & 39.2 & 49.3 & & \\
	& $1^3P_0 \; \gamma$ 
	&  55 & 128 & 79.7 & 88.6 & & \\
\hline
$2^3P_2 $ & $2^3S_1 \; \gamma$ 
	&  55  & 57.3  & 49.4 & 73.8 & & \\
	& $1^3S_1 \; \gamma$ 
	&  14 & & 19.2 & 25.8 & & \\
	& $1^3D_3 \; \gamma$ 
	&  6.8 & 1.59  & 10.9 & 17.8 & & \\
	& $1D_2' \; \gamma$ 
	&   0.7 & 0.113 & 0.5 &  & & \\
	& $1D_2 \; \gamma$ 
	&   0.6  & 0.269 & 1.5 & 3.2 \footnotemark[1] & & \\
	& $1^3D_1 \; \gamma$ 
	&   0.1 & 0.035 & 0.1  & 0.2 & & \\
$2P_1' $ & $2^3S_1 \; \gamma$ 
	&  5.5 & 9.1 & 5.9 & 5.4 & &\\
	& $2^1S_0 \; \gamma$ 
	&   52 & 72.5 & 58.0 & & &\\
	& $1^3S_1 \; \gamma$ 
	&  0.6& & 2.5 & 2.1 & & \\
	& $1^1S_0 \; \gamma$ 
	&  19 & & 20.1 & & &\\
	& $1D_2' \; \gamma$ 
	&  5.5 & 1.2 & 3.5 &  & &\\
	& $1D_2 \; \gamma$ 
	&  1.3 & 0.149 & 2.5 & 11.5 \footnotemark[1] & &\\
	& $1^3D_1 \; \gamma$ 
	&   0.2 & 0.073 & 0.3 & 0.4 & &\\
$2P_1 $ & $2^3S_1 \; \gamma$ 
	&   45 & 37.9  & 32.1 & 54.3  & & \\
	& $2^1S_0 \; \gamma$ 
	&   5.7 & 11.7 &  8.1 & & & \\
	& $1^3S_1 \; \gamma$ 
	&   5.4 & & 15.3 & 22.1 & &\\
	& $1^1S_0 \; \gamma$ 
	&   2.1 & & 3.1 & & &\\
	& $1D_2' \; \gamma$ 
	&  0.8 & 0.021 &  1.2 &  & &\\
	& $1D_2 \; \gamma$ 
	&  3.6 & 0.418 & 3.9 & 9.8 \footnotemark[1] & &\\
	& $1^3D_1 \; \gamma$ 
	&  1.6 & 0.184 & 1.6 & 0.3 & &\\
$2^3P_0 $ & $2^3S_1 \; \gamma$ 
	& 42 & 29.2 & 25.5 & 41.2  & &\\
	& $1^3S_1 \; \gamma$ 
	& 1.0 & & 16.1 & 21.9 & &\\
	& $1^3D_1 \; \gamma$ 
	&   4.2 & 0.036 & 3.2 & 6.9 & &\\
\end{tabular}
\end{ruledtabular}

\footnotetext[1]{We identify the $1D_2'$ and $1D_2$ states with the 
$^1D_2$ and $^3D_2$ states of Eichten and Quigg \cite{eichten94}. }

\end{table*}

Radiative transitions which flip spin are described by magnetic dipole
(M1) transitions.  The rates for magnetic dipole transitions in 
quarkonium bound states are given in the nonrelativistic approximation 
by \cite{JDJ,Nov78}
\begin{equation}
\Gamma( i\to f \gamma) = {{\alpha}\over 3} \; \mu^2 \; \omega^3 (2J_f +1)
 \, |\,\langle f | \; j_0(kr/2) \; |\,  i \rangle\, |^2  
\end{equation}
where 
\begin{equation}
\mu= \frac{e_c}{m_c} - \frac{e_{\bar{b}}}{m_{\bar{b}}},
\end{equation}
$e_c$ and $e_{\bar{b}}$ are the $c$-quark  and $b$-antiquark charges 
in units of $|e|$ ($e_c=2/3$ and $e_{\bar{b}}=1/3$), and $m_c$ and $m_b$ are 
the quark masses given above.

The M1 widths and overlap integrals are given in Table 
\ref{tab:m1}. 
They are compared to other calculations in Table \ref{tab:m1comp}.
Transitions in which the principle quantum number 
changes are refered to as hindered transitions which
are not allowed in the non-relativistic limit due to the orthogonality 
of the wavefunctions. 
M1 transitions, especially hindered transitions,
are notorious for their sensitivity to relativistic 
corrections \cite{m1}.   
In our calculations the wavefunction orthogonality is broken 
by including a smeared  hyperfine interaction directly in the 
Hamiltonian so that the $^3S_1$ and $^1S_0$ states have slightly 
different wavefunctions. 
Ebert {\it et al.}  are more rigorous in how they include relativistic 
corrections \cite{ebert03} but 
to improve the $J/\psi \to \eta_c \gamma$ result they modify the 
confining potential  by making it a 
linear combination of Lorentz vector and Lorentz scalar pieces.

Given the sensitivity of radiative transitions to details of the models,
precise measurements of electromagnetic transition rates would 
provide stringent tests of the various calculations.

\begin{table*}
\caption{M1 transition rates. The matrix elements were obtained using the wavefunctions 
of the GI model \cite{godfrey85}.
\label{tab:m1}}
\begin{ruledtabular}
\begin{tabular}{l l r c c c c  } 
Initial & Final & $M_i$ & $M_f$ &  $\omega$ & 
	$\langle f | \; j_0(kr/2) \; |\,  i \rangle $  & Width   \\
state & state & (MeV) & (MeV) & (MeV) & (GeV$^{-1}$) &  (keV) \\
\hline 
$1^3S_1 $ & $1^1S_0 \; \gamma$ & 6338 & 6271 & 67  & 0.995 & 0.08 \\
$2^3S_1 $ & $2^1S_0 \; \gamma$ & 6887 & 6855 & 32  & 0.992 & 0.01 \\
	  & $1^1S_0 \; \gamma$ &  & 6271 & 588  & 0.102 & 0.6 \\
$2^1S_0 $ & $1^3S_1 \; \gamma$ & 6855 & 6338 & 498  & -0.57 & 0.3 \\
$3^3S_1 $ & $3^1S_0 \; \gamma$ & 7272 & 7250 & 22  & 0.992 & 0.003 \\
	  & $2^1S_0 \; \gamma$ & 	& 6855 & 405  & 0.109 & 0.2 \\
	  & $1^1S_0 \; \gamma$ & 	& 6271 & 932  & 0.05 & 0.6 \\
$3^1S_0 $ & $2^3S_1 \; \gamma$ & 7250 & 6887 & 354  & -0.04 & 0.06 \\
	  & $1^3S_1 \; \gamma$ & 	& 6338 & 855  & 0.09 & 4.2 \\
\end{tabular}
\end{ruledtabular}
\end{table*}

\begin{table}
\caption{Comparison of M1 transition rates. Note that
no effort has been made to scale the results to 
a common phase space. 
\label{tab:m1comp}}
\begin{ruledtabular}
\begin{tabular}{l l c c c c c } 
Initial & Final &  GI  
		& EFG \cite{ebert03} 
		& GKLT \cite{gershtein95b}
		& EQ \cite{eichten94}
		& FU \cite{fulcher99} \\
state & state 	& \multicolumn{5}{c}{Widths (eV)} \\
\hline 
$1^3S_1 $ & $1^1S_0 \; \gamma$ & 80  & 33   & 60   & 134.5 & 59 \\
$2^3S_1 $ & $2^1S_0 \; \gamma$ & 10  & 17   & 10   &  28.9 & 12 \\
	  & $1^1S_0 \; \gamma$ & 600 & 428  & 98 & 123.4 & 122 \\
$2^1S_0 $ & $1^3S_1 \; \gamma$ & 300 & 488  & 96 & 93.3 & 139 \\
\end{tabular}
\end{ruledtabular}
\end{table}

\section{Hadronic Transitions}

Hadronic transitions between quarkonium levels are needed to
estimate branching ratios and discuss search strategies for 
these states. In fact, these are the dominant decays for both the 
$\psi(2S)$ and $\Upsilon(2S)$ states.
There have been numerous theoretical estimates of 
hadronic transitions over the years 
\cite{Yan80,Kua81,Kua88,Kua90,Rosner03,Vol80,Nov81,Iof80,Vol86,Vol03a,Ko94,Mox88,Ko93}.  
In some cases the estimates disagree by orders of 
magnitude \cite{Rosner03}.  
Hadronic transitions are typically described as a two-step 
process in which the gluons are first emitted from the heavy quarks 
and then recombine into light quarks.  A multipole expansion of the 
colour gauge field is employed to describe the emission process where 
the intermediate colour octet quarkonium state is modeled by some sort 
of quarkonium hybrid wavefunction.  However, the main disagreement 
between predictions arises from how the rehadronization step is estimated.  
To some extent this latter uncertainty can be reduced by employing 
the multipole expansion of the colour gauge fields 
developed by Yan and collaborators \cite{Yan80,Kua81,Kua88,Kua90}
together with the Wigner-Eckart theorem 
to estimate the E1-E1 transition rates \cite{Yan80}  and fixing the 
reduced matrix elements by rescaling measured transition rates.  When 
no measured transitions exist we instead rescale the theoretical estimates of 
related matrix elements \cite{Kua81}.  This is 
the approach used by Eichten and Quigg \cite{eichten94}. 
In addition to E1-E1 transitions there will be other transitions such 
as $^3S_1\to ^3S_1 +\eta$ which goes via M1-M1 \& E1-M2 multipoles
and spin-flip transitions such as $^3S_1 \to ^1P_1 \pi\pi$ which
goes via E1-M1 \cite{Kua81}.  
These transitions are suppressed by inverse powers 
of the quark masses and are expected to be small compared to the E1-E1 
and electromagnetic transitions. The $2^3S_1\to 1^3S_1 
+\eta$ transitions are further suppressed due to being at the limit of 
available phase space.

The differential rate for E1-E1 transitions from an initial quarkonium 
state $\Phi'$ to the final quarkonium state $\Phi$, and a system of 
light hadrons, $h$, is given by the expression \cite{Yan80,Kua81}:
\begin{equation}
{{d\Gamma}\over {d{\cal M}^2}} [\Phi' \to \Phi +h] 
= (2J+1)\sum_{k=0}^2
\left\{ { 
\begin{array}{ccc}  k & \ell ' & \ell \\ s & J & J'
\end{array} 
} \right\}
^2 A_k(\ell' , \, \ell )
\end{equation}
where $\ell'$, $\ell$ are the orbital angular momentum  and
$J'$, $J$ are the total angular momentum of the 
initial and final states respectively,  $s$ is the spin of the 
$Q\bar{Q}$ pair, ${\cal M}^2$ is the invariant mass squared of 
the light hadron system, $\{ {\cdots \atop \cdots} \}$ is a $6-j$ symbol, 
and $A_k(\ell' , \, \ell )$ are the reduced matrix elements. The 
magnitudes of the $A_k(\ell' , \, \ell )$ are model dependent with a 
large variation in their estimates.  In the soft-pion limit the 
$A_1$ contributions are suppressed so, as is the usual practice, we 
will take $A_1(\ell' , \, \ell )=0$.  For the remaining rates we use 
scaling arguments taking measured rates as input 
or, when no measured rates exist,
we rescale the rates predicted for
the $b\bar{b}$ system by Kuang and Yan \cite{Kua81} 
to obtain the $c\bar{b}$ rates.  
The amplitudes for E1-E1 transitions depend 
quadratically on the interquark separation so the scaling law between 
a $c\bar{b}$ rate and the corresponding $Q\bar{Q}$ rate is given by 
\cite{Yan80}
\begin{equation}
{{\Gamma(c\bar{b})}\over {\Gamma(Q\bar{Q})}}=
{{ \langle r^2 (c\bar{b}) \rangle^2}\over 
{ \langle r^2 (Q\bar{Q}) \rangle^2}}
\end{equation}
up to phase space.
The scaling factors used to relate the input rates to the $c\bar{b}$ 
rates are given in Table \ref{tab:had-me}.

There is a large variation 
in the reduced rates.  For example, for the transition $1^3D_1\to 
1^3S_1 +\pi\pi$ estimates for $A_2(2,0)$ differ by almost three orders 
of magnitude \cite{Rosner03,Kua81,Mox88,Ko93}.  
We point this out as a cautionary note to the reader.
The reduced rates are summarized in Table \ref{tab:had-me}. 
For the $2^3S_1\to 1^3S_1 +\pi\pi$ reduced rate we take an average of 
the results taken from rescaling the $\psi'\to J/\psi +\pi\pi$ and 
$\Upsilon (2S) \to \Upsilon +\pi\pi$ rates.  The $3^3S_1\to 2^3S_1 
+\pi\pi$ and $3^3S_1\to 2^3S_1 +\pi\pi$ values were obtained by 
rescaling the corresponding $b\bar{b}$ transitions \cite{pdg02}.  
$A_0(1,1)$ and $A_2(1,1)$, corresponding to
 $2^3P_0\to 1^3P_0 +\pi\pi$ and  $2^3P_2\to 1^3P_1 +\pi\pi$ 
respectively, were obtained by rescaling the appropriate $b\bar{b}$
rate predictions given by Kuang and Yan \cite{Kua81}.  
As pointed out above 
there is considerable variation in the $A_2(2,0)$ amplitude needed for 
the $1^3D_J\to 1^3S_1 +\pi\pi$ transitions.  The largest predicted rate
for the $\Upsilon(1D)\to \Upsilon(1S) +\pi\pi$ transition comes from
Kuang and Yan \cite{Kua81} and has been ruled 
out by a recent CLEO limit \cite{Bonvicini:2004yj}. The CLEO limit
is about a 
factor of 3 larger than the rate predicted by Ko \cite{Ko93}.  The 
reduced rate for the $c\bar{b}$ system found by rescaling the recent
BES measurement \cite{Bai03} of the $\psi''\to 
J/\psi \pi\pi$ rate  is considerably larger than the rate found by 
rescaling the $\Upsilon(1D)\to \Upsilon(1S) +\pi\pi$ CLEO limit.  
However, it is likely that one can reconcile the $b\bar{b}$ and 
$c\bar{c}$ results by properly taking into account $2^3S_1-1^3D_1$ 
mixing \cite{Kua90,Liu04,Rosner04}. We will therefore assume a reduced 
rate  of $A_2(2,0)=21$~keV 
which is based on the CLEO limit on the transition
$\Upsilon(1D)\to \Upsilon(1S) +\pi\pi$ \cite{Bonvicini:2004yj}.  
The $1D\to 1S \pi\pi$ transitions is the subject of recent interest
\cite{Rosner03,Liu04,Rosner04} and
as the experimental measurements improve it would be 
useful to revisit these calculations. 
The uncertainty in these hadronic transitions could easily lead 
to factors of two errors in the resulting branching ratios.
A final note is that we have not considered coupled channel effects to 
$D\bar{D}$ and $B\bar{B}$ for the $c\bar{c}$ and $b\bar{b}$ states 
respectively which could make a considerable contribution to states 
close to threshold \cite{Rosner04}.

The reduced rates of Table \ref{tab:had-me} 
are used to obtain the $c\bar{b}$ 
hadronic transitions which are summarized in Table \ref{tab:hadtran}.
We do not include decays of the type 
$2^{3,1}P_J\to 1^{3,1}P_{J'}$, as they are expected to be 
small compared to the decays we included.  Likewise, transitions 
with $\eta$ and $\pi^0$ in the final state are possible but are 
expected to have much smaller partial widths
and $\eta$ transitions are further suppressed by phase space.
Although
the $3^3S_1$ and $3^1S_0$ states are expected to be above $BD$ 
threshold, and therefore relatively broad, we include the 
two pion transitions for the sake of completeness.

\begin{table*}
\caption{Estimates of reduced rates for E1-E1 hadronic 
transitions between $c\bar{b}$ levels.  
\label{tab:had-me}}
\begin{ruledtabular}
\begin{tabular}{l l c c} 
Transition & $(Q\bar{Q}):$   rate (keV) 
	& $\langle r^2(c\bar{b})\rangle / \langle r^2(Q\bar{Q})\rangle$
	& Reduced $c\bar{b}$ rate (keV) \\
\hline 
$2^3S_1 \to 1^3S_1 + \pi\pi$ & $(c\bar{c})$: $146\pm 14$ \footnotemark[1]
		& 0.75 & $A_0(0,0)=82 \pm 8$	\\
			& $(b\bar{b})$:  $12.2\pm 2$ \footnotemark[1]
		& 1.63 & $A_0(0,0)=33\pm 5$	\\
		& Average	&	& $57 \pm 7$ \\
$3^3S_1 \to 2^3S_1 + \pi\pi$ & $(b\bar{b})$: $1.26\pm 0.25$ \footnotemark[1]
		& 1.56 & $A_0'(0,0)=3.1 \pm 0.6$	\\
$3^3S_1 \to 1^3S_1 + \pi\pi$ & $(b\bar{b})$: $1.72\pm 0.25$ \footnotemark[1]
		& 1.56 & $A_0''(0,0)=4.2 \pm 0.6$	\\
$2^3P_0 \to 1^3P_0 + \pi\pi$ & $(b\bar{b})$:  0.4 \footnotemark[2]
		& 1.56 & $A_0(1,1)=2.92$	\\
$2^3P_2 \to 1^3P_1 + \pi\pi$ & $(b\bar{b})$:  0.01 \footnotemark[2]
		& 1.57 & $A_2(1,1)=0.164$	\\
$1^3D_1 \to 1^3S_1 + \pi\pi$ & $(c\bar{c})$:  $120\pm 57$ \footnotemark[3]
		& 0.78 & $A_2(2,0)=360\pm 170$	\\
		& $(c\bar{c})$:  $<92$ 90\% C.L. \footnotemark[4]
		& 0.78 & $A_2(2,0) < 280$	\\
		& $(b\bar{b})$:  24 \footnotemark[5]
		& 1.6 & $A_2(2,0) =307$	\\
		& $(b\bar{b})$:  0.07 \footnotemark[6]
		& 1.6 & $A_2(2,0) =0.9$	\\
		& $(b\bar{b})$:  $0.56\pm0.07$ \footnotemark[7]
		& 1.6 & $A_2(2,0) =7.2$	\\
\end{tabular}
\end{ruledtabular}
\footnotetext[1]{From PDG Ref.\cite{pdg02}.}
\footnotetext[2]{From Kuang and Yan using model C: modified Richardson 
potential Ref. \cite{Kua81}.}
\footnotetext[3]{From BES Ref.\cite{Bai03}.}
\footnotetext[4]{CLEO 90\% C.L. upper limit \cite{Skwarnicki:2003wn}.}
\footnotetext[5]{From Kuang and Yan using model A: linear plus Coulomb
potential Ref. \cite{Kua81}.}
\footnotetext[6]{From Moxhay Ref. \cite{Mox88}.}
\footnotetext[7]{From BR prediction of Ko \cite{Ko93} combined with the total 
width predicted by Kuang and Rosner \cite{Kwo88}.}

\end{table*}

\begin{table*}
\caption{Rates for two-pion E1-E1 hadronic transitions.  The reduced 
rates are denoted by $A_k(\ell',\, \ell)$ where $k$ is the rank of the 
irreducible tensor and $\ell'$ and $\ell$ are the orbital anular 
momenta of the initial and final states.
\label{tab:hadtran}}
\begin{ruledtabular}
\begin{tabular}{l c c} 
Transition & Expression for Rate &  $(c\bar{b})$ rate (keV)  \\
\hline 
$2^3S_1 \to 1^3S_1 + \pi\pi$ & $A_0(0,\, 0)$ & $57\pm 7$ \\
$2^1S_0 \to 1^1S_0 + \pi\pi$ & $A_0(0,\, 0)$ & $57\pm 7$ \\
$3^3S_1 \to 2^3S_1 + \pi\pi$ & $A_0'(0,\, 0)$ & $3.1\pm 0.6$  \\
$3^1S_0 \to 2^1S_0 + \pi\pi$ & $A_0'(0,\, 0)$ & $3.1\pm 0.6$ \\
$3^3S_1 \to 1^3S_1 + \pi\pi$ & $A_0''(0,\, 0)$ & $4.2\pm 0.6$ \\
$3^1S_0 \to 1^1S_0 + \pi\pi$ & $A_0''(0,\, 0)$ & $4.2\pm 0.6$ \\
\hline
$2^3P_2 \to 1^3P_2 + \pi\pi$ 
	& $\frac{1}{3}A_0(1,\, 1) + \frac{1}{4}A_1(1,\, 1) 
		+\frac{7}{60}A_2(1,\, 1) $ & 1.0 \\
$2^3P_2 \to 1P_1' + \pi\pi$ 
	& $ \frac{1}{12}A_1(1,\, 1) +\frac{3}{20}A_2(1,\, 1) $
			\footnotemark[1] & 0.004 \footnotemark[2]\\
$2^3P_2 \to 1P_1 + \pi\pi$ 
	& $ \frac{1}{12}A_1(1,\, 1) +\frac{3}{20}A_2(1,\, 1) $ 
			\footnotemark[1] & 0.021 \footnotemark[2]\\
$2^3P_2 \to 1^3P_0 + \pi\pi$ 
	& $ \frac{1}{15}A_2(1,\, 1) $ &  0.011 \\
$2P_1' \to 1^3P_2 + \pi\pi$ 
	& $\frac{5}{36}A_1(1,\, 1)+\frac{1}{4}A_2(1,\, 1) $ 
			\footnotemark[3] & 0.004 \footnotemark[2]\\
$2P_1 \to 1^3P_2 + \pi\pi$ 
	& $\frac{5}{36}A_1(1,\, 1)+\frac{1}{4}A_2(1,\, 1) $ 
			\footnotemark[3] & 0.037 \footnotemark[2]\\
$2P_1' \to 1P_1' + \pi\pi$ 
	& $ A_0(1,\, 1)+A_1(1,\, 1) +\frac{1}{3}A_2(1,\, 1) $
			\footnotemark[4]  & 1.2 \footnotemark[5] \\
$2P_1' \to 1P_1 + \pi\pi$ 
	& $ A_0(1,\, 1)+A_1(1,\, 1) +\frac{1}{3}A_2(1,\, 1) $
			\footnotemark[4]  & 0.1 \footnotemark[5] \\
$2P_1 \to 1P_1' + \pi\pi$ 
	& $\frac{1}{3}A_0(1,\, 1) + \frac{1}{12}A_1(1,\, 1) 
		+\frac{1}{12}A_2(1,\, 1) $ \footnotemark[6]
			&  0.02 \footnotemark[5]\\
$2P_1 \to 1P_1 + \pi\pi$ 
	& $\frac{1}{3}A_0(1,\, 1) + \frac{1}{12}A_1(1,\, 1) 
		+\frac{1}{12}A_2(1,\, 1) $ \footnotemark[6]
			&  2.7 \footnotemark[5]\\
$2P_1' \to 1^3P_0 + \pi\pi$ & $ \frac{1}{9}A_1(1,\, 1) $ 
			\footnotemark[7] & 0 \\
$2P_1 \to 1^3P_0 + \pi\pi$ & $ \frac{1}{9}A_1(1,\, 1) $ 
			\footnotemark[7] & 0 \\
$2^3P_0 \to 1^3P_2 + \pi\pi$ 
	& $\frac{1}{3}A_2(1,\, 1) $ & 0.0547 \\
$2^3P_0 \to 1P_1' + \pi\pi$ & $ \frac{1}{3}A_1(1,\, 1) $ 
			\footnotemark[8] & 0 \\
$2^3P_0 \to 1P_1 + \pi\pi$ & $ \frac{1}{3}A_1(1,\, 1) $ 
			\footnotemark[8] & 0 \\
$2^3P_0 \to 1^3P_0 + \pi\pi$ & $ \frac{1}{3}A_0(1,\, 1) $ & 0.97 \\
\hline
$1^3D_{1,3} \to 1^3S_1 + \pi\pi$ & $\frac{1}{5}A_2(2,\, 0)$ \footnotemark[9]
			& 4.3 \\
$1D_2' \to 1^3S_1 + \pi\pi$ & $\frac{1}{5}A_2(2,\, 0)$ \footnotemark[9]
			& 2.1 \footnotemark[2]\\
$1D_2 \to 1^3S_1 + \pi\pi$ & $\frac{1}{5}A_2(2,\, 0)$ \footnotemark[9]
			& 2.2 \footnotemark[2]\\
$1D_2' \to 1^1S_0 + \pi\pi$ & $\frac{1}{5}A_2(2,\, 0)$ \footnotemark[9]
			& 2.2 \footnotemark[2]\\
$1D_2 \to 1^1S_0 + \pi\pi$ & $\frac{1}{5}A_2(2,\, 0)$ \footnotemark[9]
			& 2.1 \footnotemark[2]\\
\end{tabular}
\end{ruledtabular}
\footnotetext[1]{The expression is for the $^3P_2\to^3P_1$ transition.}
\footnotetext[2]{These rates include the appropriate mixing angles
defined in eqn. and given in Table \ref{tab:masses}.}
\footnotetext[3]{The expression is for the $^3P_1\to^3P_2$ transition.}
\footnotetext[4]{The expression is for the $^1P_1\to^1P_1$ transition.}
\footnotetext[5]{These rates include the appropriate mixing angles.
We assume a +ve phase between the $^1P_1\to^1P_1$ 
and $^3P_1\to^3P_1$ amplitudes. The $2P_1'\to 1P_1$ and $2P_1\to 1P_1'$
widths are most sensitive to this phase.  In any case the widths are 
expected to be quite small and not particularly important.}
\footnotetext[6]{The expression is for the $^3P_1\to^3P_1$ transition.}
\footnotetext[7]{The expression is for the $^3P_1\to^3P_0$ transition.}
\footnotetext[8]{The expression is for the $^3P_0\to^3P_1$ transition.}
\footnotetext[9]{These correspond to the $^3D_J\to ^3S_1$ or $^1D_2\to 
^1S_0$ transitions as appropriate.}

\end{table*}

\section{Weak Decays}

The final ingredient needed in a study of $B_c$ phenomenology is the 
$B_c$ width and its weak decay partial widths.  The details of 
$B_c$ decay have been given elsewhere 
\cite{kiselev03a,kiselev03,Jones:1998ub,ivanov00,chang94,nobes00,ebert03b,chang01,Du:ws,Anisimov:1998xv}. 
For completeness we give a brief overview of the essential features of 
these decays and  summarize the weak decay branching ratios in 
Table \ref{tab:weak}.  We refer the interested reader to the original 
literature for details of the calculations
\cite{Capstick:1989ra,kiselev03a,kiselev03,Jones:1998ub,ivanov00,chang94,nobes00,ebert03b,chang01,Du:ws,Anisimov:1998xv}.

For a rough estimate of the $B_c$ width we can treat the 
$\bar{b}$-quark and $c$-quark decay independently so that
$B_c$ decay can be divided into three classes: (i) the $\bar{b}$-quark 
decay with spectator $c$-quark, (ii) the $c$-quark 
decay with spectator $\bar{b}$-quark, and (iii) the annihilation 
$B_c^+\to \ell^+ \nu_\ell \; (c\bar{s}, \; u\bar{s})$, where 
$\ell=e,\; \mu, \; \tau$.  The total width is the sum over partial 
widths
\begin{equation}
\Gamma(B_c \to X)=\Gamma(b\to X) + \Gamma(c\to X) +\Gamma(ann)
\end{equation}
In addition there is a Pauli interference contribution to the 
$\bar{b}\to \bar{c}c\bar{s}$ decay from the $c$-quark spectator
which we ignore in this crude estimate.

In the spectator approximation:
\begin{equation}
\Gamma (\bar{b} \to X) = {{9G_F^2 |V_{cb}|^2 m_b^5}\over {192\pi^3}}
\simeq 4.8\times 10^{-4}\hbox{ eV}
\end{equation}
and
\begin{equation}
\Gamma (c \to X) = {{5G_F^2 |V_{cs}|^2 m_c^5}\over {192\pi^3}} 
\simeq 3.3\times 10^{-4}\hbox{ eV}
\end{equation}
where we used $|V_{cb}|=0.0412$, $|V_{cs}|=0.974$, 
$m_b=4.25$~GeV, and $m_c=1.25$~GeV \cite{pdg02}.

Annihilation widths such as $c\bar{b}\to \ell \nu_\ell$
are given by the expression
\begin{equation}
\Gamma = {{G_F^2}\over {8\pi}} |V_{bc}|^2 f_{B_c}^2 M_{B_c} \sum_i 
m_i^2 \left( { 1- {{m_i^2}\over{M_{B_c}^2} } } \right)^2 C_i
\end{equation}
where $m_i$ is the mass of the heavier fermion in the given decay 
channel.  For lepton channels $C_i=1$ while for quark channels 
$C_i=3|V_{qq'}|^2$. 
The pseudoscalar decay constant, $f_{B_c}$, is  defined by:
\begin{equation}
\langle 0 | \bar{b}(x)\gamma^\mu \gamma_5 \, c(x) 
| B_c(k) \rangle = i f_{B_c} \, V_{cb} \, k^\mu
\end{equation}
where 
$V_{cb}$ is the $cb$ element of the 
Cabibbo-Kobayashi-Maskawa matrix, and $k^\mu$ is the four-momentum of 
the $B_c$ meson.  In the non-relativistic limit the pseudoscalar 
decay constant is proportional to the wavefunction at the origin and 
is given by the van Royen-Weisskopf formula
\begin{equation}
f_{B_c} = {{2\sqrt{3}}\over M} \psi(0).
\end{equation}
This result is modified by QCD corrections and
relativistic effects, which are included using the Mock-Meson 
approach  or other relativistic quark models.  
The predictions of the various calculations including Lattice QCD 
are summarized in Table \ref{tab:fp}.  Using the mock-meson result of 
$f_{B_c}=410$~MeV \cite{Capstick:1989ra}, which is consistent with most 
other predictions, leads to the annihilation width of
\begin{equation}
\Gamma(ann)=67\times 10^{-6} \hbox{ eV}
\end{equation}

Adding this result to the spectator contributions 
gives $\Gamma(total)=8.8\times 10^{-4}$~eV corresponding to 
a $B_c$ lifetime of $\tau=0.75$~ps which is 
in rough agreement with the measured 
value of $\tau=0.46^{+0.18}_{-0.16}$~ps.  A more careful calculation 
by Kiselev gives $\tau=0.5$~ps \cite{kiselev03a}.
The approximate branching fractions for the
$b$-decay, $c$-decay, and annihilation processes are
54\%, 38\%, and 8\% respectively.
 These BR's are modified by strong interaction 
effects \cite{gershtein95,kiselev03a,ivanov00,chang94,nobes00,chang01} 
which are included in some recent calculations of
BR's. The branching ratios of some
prominent decay modes are summarized in Table \ref{tab:weak}.
In addition, weak $B_c$ decays to
P-wave charmonium states, $\chi_c$ or $h_c$,
are potentially important decay modes \cite{chang01} but we will 
neglect them in favour of simpler to observe decay chains.

\begin{table*}
\caption{Comparison of predictions for  the pseudoscalar decay 
constant of the $B_c$ meson.
\label{tab:fp}}
\begin{ruledtabular}
\begin{tabular}{c c c c c c c c} 
 GI \cite{Capstick:1989ra}  
	& EFG \cite{ebert03} 
	& GKLT \cite{gershtein95b} 
	& EQ \cite{eichten94}
	& Fu \cite{fulcher99}
	& Ki\cite{kiselev03}
	& Lattice \cite{davies96} 
	& Lattice \cite{Jones:1998ub}  \\
\hline 
$410\pm 40$	& 433  & $500\pm 80$ & 500 \footnotemark[1] 
		& 517 & $395\pm 15$ & $440\pm 20$
		& $420\pm 13$ \\
\end{tabular}
\end{ruledtabular}

\footnotetext[1]{Using Buchm\"uller-Tye potential.}

\end{table*}

\begin{table}
\caption{Branching ratios in \% 
for some prominent exclusive $B_c^+$ decays.
\label{tab:weak}}
\begin{center}
\begin{tabular}{l l l l l l l}  \hline \hline
     & Mode            	& Kis \cite{kiselev03a} 
			& IKS \cite{ivanov00}
			& CC \footnotemark[1] \cite{chang94} 
			& EFG \cite{ebert03b}
			& NW \cite{nobes00}\\ 
\hline
$B_c^+$	& $\to J/\psi e^+ \nu $ & 1.9	& 2.3  & 2.4 & 1.2 & 1.5 \\
	& $\to \eta_c e^+ \nu $ & 0.75	& 0.98	& 1.0 & 0.42 & 0.52\\
	& $\to B^0 e^+ \nu $ 	& 0.34	& 0.15 & 0.16 & 0.042 & 0.05 \\
	& $\to B^{*0} e^+ \nu $ & 0.58 	& 0.16 & 0.23 & 0.12 & 0.05\\
	& $\to B_s^0 e^+ \nu $ 	& 4.03	& 2.0  & 1.9  & 0.84 & 0.94 \\
	& $\to B_s^{*0} e^+ \nu $ & 5.06 & 2.6 & 3.1  & 1.75 & 1.44\\
	& $\to J/\psi \pi^+ $ 	& 0.13	&	& 0.22 & 0.061 &  \\
	& $\to J/\psi \rho^+ $ 	& 0.40	&  	& 0.66 & 0.16 & \\
	& $\to \eta_c \pi^+$	& 0.20	&	& 0.23 & 0.085 & \\
	& $\to \eta_c \rho^+$	& 0.42	&	& 0.61 & 0.21 & \\
	& $\to B_s^0 \pi^+ $ 	& 16.4	&	& 5.1  & 2.52 & \\
	& $\to B_s^0 \rho^+ $ 	& 7.2	&	& 3.9  & 1.41 & \\
	& $\to B_s^{*0} \pi^+ $ & 6.5	&	& 4.5  & 1.61 & \\
	& $\to B_s^{*0} \rho^+ $& 20.2	&	& 13.1 & 11.1 & \\
	& $\to B_s^{0} K^+ $ 	& 1.06	&	& 0.37 & 0.21 & \\
	& $\to B_s^{*0} K^+ $ 	& 0.37	&	& 0.26 & 0.11 & \\
	& $\to B^0 \pi^+ $ 	& 1.06	&	& 0.29 & 0.10 & \\
	& $\to B^0 \rho^+ $ 	& 0.96	&	& 0.52 & 0.13 & \\
	& $\to B^{*0} \pi^+ $ 	& 0.95	&	& 0.25 & 0.03 & \\
	& $\to B^{*0} \rho^+ $ 	& 2.57	&	& 1.0  & 0.68 & \\
	& $\to B^+ \bar{K}^0 $ 	& 1.98	&	& 0.30 & 0.24 & \\
	& $\to B^+ \bar{K}^{*0} $ & 0.43 &	& 0.21 & 0.09 & \\
	& $\to B^{*+} \bar{K}^{0} $ & 1.60 &	& 0.23 & 0.11 & \\
	& $\to B^{*+} \bar{K}^{*0} $ & 1.67 &	& 0.44 & 0.84 & \\
	& $\to \tau^+ \nu_\tau $ & 1.6 &	& 	& & \\
	& $\to c\bar{s} $ 	& 4.9 &		& 	& & \\
\hline \hline
\end{tabular}
\footnotetext[1]{Using the PDG \cite{pdg02}
central value for the total width of $\Gamma= 1.43\times 10^{-3}$~eV
to obtain the BR's.}
\end{center}

\end{table}

\section{Experimental Signatures and Search Strategies}

$B_c$ mesons offer a rich 
spectroscopy of narrow states to study;  
there are 2 $S$-wave, 2 $P$-wave, and  1 $D$-wave $B_c$ multiplets below
$BD$ threshold.  Because $B_c$ mesons carry flavour they cannot 
annihilate into gluons and are expected to be quite narrow; $<100$~keV.  
In addition, the $F$-wave states are just above threshold 
so might also be relatively narrow due to the angular momentum barrier 
which would suppress the decay \cite{strongwidth}.  
Two ingredients are necessary for the study of $B_c$ spectroscopy; 
that they be produced in sufficient quantity and that 
they yield a signal that can be distinguished from background.

$B_c$ production production proceeds via the hard associative 
production of the two heavy quark pairs $c\bar{c}$ and $b\bar{b}$ 
which suppresses the $B_c$ yield relative to beauty hadrons by 
${\cal O}(10^{-3})$ \cite{cheung93}.  Because 
fragmentation dominates at high-$p_T$ it has proven useful to 
describe $B_c$ meson production by hadronization of 
individual high-$p_T$ partons 
using the factorization formalism based on non-relativistic 
QCD \cite{braaten96}. 
This approach was utilized by a number of 
authors to calculate $B_c$ production at hadron colliders.   
At the Tevatron, with acceptance cuts of $p_t>6$~GeV and $|y|<1$, 
Cheung estimates that 
${\cal O} (10^7)$ $B_c$ mesons should be produced for 
an integrated luminosity of 1~fb$^{-1}$ \cite{cheung96}.
At the LHC ${\cal O} (10^9)$ $B_c$ mesons are expected to be produced
for 100~fb$^{-1}$ with the kinematic cuts of $p_t>10$~GeV and $|y(B_c)|<2.5$
\cite{cheung96}.
The $B_c^*$ cross sections are expected to be 50-100\% larger than the 
$B_c$ cross sections \cite{chang96,cheung93,cheung96}.
The $p_T$ distribution falls rapidly so that the small $p_T$ 
region is quite important \cite{chang96} but also making the cross 
sections sensitive to the exact values of the kinematic cuts
\cite{chang03,Chang:1994aw}.
With the high luminosity of the LHC one expects a sizable number 
of $P$- and $D$-wave $(c\bar{b})$ states ($\sim 20\%$ and $\sim 2\%$ 
of the total inclusive $B_c$ cross section respectively)
as well as excited $S$-wave 
states to be produced ($2S/1S \sim 0.6$)
\cite{cheung96,Cheung:1995ir,chang93}.  
The LHC should therefore produce sufficient $B_c$ mesons to allow the 
study of the $c\bar{b}$ spectroscopy and decays.  We will use these 
numbers as the starting point to estimate the number of $B_c$'s 
produced in a particular decay chain.

Irregardless of which state is produced, it will eventually cascade
decay to the $B_c$ ground state via electromagnetic and hadronic 
transitions  so that the $B_c$ must be observed in order to 
reconstruct the parent particle for the particular decay chain.  
Prominent decays of the $B_c$ are given in Table \ref{tab:weak}. 
$B_c$ decays with a $J/\psi$ in the final state
such as $J/\psi +X$ where $X$ can be a $\pi^+$, $\rho^+$,  
or $\ell ^+ \nu_\ell$ 
are especially useful 
as the $J/\psi\to \ell^+\ell^-$ provides a useful trigger for $B_c$ 
events.
The golden channel to detect $B_c$ is $B_c \to J/\psi \pi (\rho)$ but 
their BR's are quite small ${\cal O}(0.2-0.4\%)$ resulting in a 
combined BR for
$B_c\to J/\psi + \pi^+ \to \ell'^-\ell'^+ \pi^+$ of $\sim 0.02\%$. 
This would yield about 2000 events for 1~fb$^{-1}$ 
integrated luminosity at the Tevatron  and about $2\times 10^5$ events 
for 100~fb$^{-1}$ integrated luminosity at the LHC. 
The decay 
$B_c^\pm \to J/\psi \ell^\pm \nu_\ell$ with $J/\psi\to e^+e^-$ has a 
distinctive signature of 3 charged leptons coming from a common 
secondary vertex and has a BR of about 2\% resulting in a 
combined BR for $B_c\to J/\psi \ell^+ \nu_\ell \to \ell'^-\ell'^+ 
\ell^+ \nu_\ell$  ($\ell=e, \, \mu$) of about 0.2\% yielding  
$2\times 10^4$ events at the Tevatron and $2\times 10^6$ at the LHC.
The trade off is that the semileptonic 
decay mode $B_c \to J/\psi + \ell \nu_\ell$ has a larger branching 
ratio but also has missing energy while 
$B_c^\pm\to J/\psi \pi^\pm$ has a smaller BR  but has the advantage that the 
$B_c$ can be fully reconstructed.  In these $B_c$ decays the $b$ quark 
decays to charm.  $B_c$ decays in which the $c$ quark decays,
such as $B_c^+\to B_s^0 \pi^+$,
have much larger BR's  and could also prove to be important modes 
if $B_c$'s can be reconstructed in these channels.

To estimate event rates we also need to include detection 
efficiencies.  Simulations by D0 and CDF \cite{Anikeev:2001rk} find 
efficiencies for the exclusive decays $B_c\to J/\psi \pi \to 
\mu^+\mu^-\pi$ and $B_c\to J/\psi \ell \nu_\ell \to \mu^+\mu^-\ell 
\nu_\ell$ of $\sim 2\%$ and $\sim 4\%$ respectively.  

In Table \ref{tab:br1} we combine the electromagnectic transitions 
widths with the hadronic transition widths to give total widths and 
BR's.  The are used 
in Table \ref{tab:events} to give estimates for the number of events 
expected at the Tevatron and LHC
for the more prominent decay chains of $B_c$ excited states.  
We assume that the ground $B_c$ state is observed in the 
$B_c\to J/\psi \ell^+ \nu_\ell \to \ell'^-\ell'^+ \ell^+ \nu_\ell$ 
and $B_c\to J/\psi + \pi^+ \to \ell'^-\ell'^+ \pi^+$ decay modes with 
BR's 2\% and 0.2\% respectively and detection efficiencies of 
$\sim 2\%$ and $\sim 4\%$ respectively.  We include a factor of 2 to 
take into account both the $e^+e^-$ and $\mu^+\mu^-$ decay modes of 
the $J/\psi$ and a factor of 2 to take into account the production of 
both charge conjugate $B_c$ states.  
To take into account the relative 
production rates of different excited states we use the production 
rates given above of  $10^7$ $B_c$'s at the Tevatron and $10^9$ 
$B_c$'s at the LHC with relative  numbers of $\times 2$ for the 
$B_c^*$, 0.2 for the $1P$ states, 0.02 for the $1D$ states, and the 
0.6 for the $2S$ relative to the $1S$ states. For the $2P$ states we 
use the same factor of 0.6 for $2P$ relative to $1P$ but this is a 
rather arbitrary assumption.  As noted already, the cross sections are 
very sensitive to the kinematic cuts so the number of events expected 
should only be taken as rough estimates.

The signal for excited $B_c$ states is a photon or pions in 
coincidence with $B_c$ decay.  
A serious omission from the estimates given in Table \ref{tab:events} 
is the neglect of tagging efficiencies for the 
photons and pions in the transitions.  The photon ID can be relatively 
high depending on the kinematics.  
For the pions, the combinatorial background is large 
and $\pi/K$ separation is not so good so one really needs to do 
studies of specific processes.  Thus,  
understanding photon and pion identification 
requires a detailed simulation study which is beyond the scope of this 
paper.  However, one might be optimistic, given the success of CDF in 
studying $\chi_c$ production \cite{Abe:1997yz} and the observation of 
the $X(3872)\to J/\psi \pi^+\pi^-$ \cite{Acosta:2003zx}
in $\bar{p}p$ collisions at the Tevatron.

Notwithstanding the previous caveat, it should 
be possible to observe the $1S$, $2S$, and $1P$ states at the 
Tevatron.  It is also possible, although only marginally so, that some 
of the $1D$ and $2P$ might also be seen.  With the higher statistics 
available at the LHC, all $c\bar{b}$ states below threshold could 
potentially be observed, although the larger backgrounds will make 
this quite challenging.  

The fact that the $B_c$ is not an eigenstate of charge conjugation 
helps simplify the search for states such as the $1P_1'$, $1P_1$, 
$1D_2'$, and $1D_2$.  The singlet component of these states allows E1 
or hadronic transitions directly to the ground state $B_c$ with
large BR's. 
This should simplify the reconstruction efforts significantly.
For example $1P_1'$ production and decay to the $B_c$ with 
its subsequent decay to $J/\psi \pi$ and $J/\psi \ell\nu_\ell$ should 
produce ${\cal O}(600)$ events in Run II at the Tevatron.  Likewise, 
$2P_1'$ and its subsequent decay should produce ${\cal O}(100)$ 
events.  These yields would be enhanced if other $B_c$ decay modes with 
larger BR's could be utilized.  
The discovery of the $2P_1'$ would yield important 
spectroscopic information in addition to being an experimental tour de 
force.  Of course, these BR's are highly sensitive to the $1P_1'-1P_1$ 
mixing angle.  With enough measurements these details can be 
constrained and different models can be differentiated.
There will also be large number of events for decay chains going via an 
intermediate $B_c^*$ such as 
$2^3S_1\stackrel{\pi\pi}{\to}1^3S_1\stackrel{\gamma}{\to}B_c$
and 
$1^3P_2\stackrel{\gamma}{\to}1^3S_1\stackrel{\gamma}{\to}B_c$.
However, it will be crucial to detect the 67~keV photon 
in $B_c^*\to B_c + \gamma$ in these cases, a very challenging 
experimental task.

\begin{table}
\caption{Partial widths and branching fractions
for strong and electromagnetic transitions. 
Details of the calculations are given in the text.
\label{tab:br1}}
\begin{center}
\begin{tabular}{l l l r } \hline \hline
Initial \phantom{www}        & Final            	& Width  	& B.F. \\
 state          & state            	& (keV)  	& \ \ (\%)  \\ 
\hline \hline
$1^3S_1$	& $1^1S_0$	   	& 0.08	 	& 100	\\
\hline
\hline
$1^3P_2$	& $1^3S_1$		& 83	& 100	\\ \hline
$1P_1'$		& $1^3S_1$		& 11	& 12.1	\\
		& $1^1S_0$		& 80	& 87.9	\\
		& Total			& 91	& 100	\\ \hline
$1P_1$		& $1^3S_1$		& 60	& 82.2	\\
		& $1^1S_0$		& 13	& 17.8	\\
		& Total			& 73	& 100	\\ \hline
$1^3P_0$	& $1^3S_1$		& 55	& 100	\\ \hline
\hline
$2^1S_0$	& $1^1S_0 +\pi\pi$	& $57\pm 7$ & 88.1 \\
		& $1P_1' +\gamma $ 	& 6.1	& 9.4 \\
		& $1P_1 +\gamma $ 	& 1.3	& 2.0 \\
		& $1^3S_1 + \gamma$ 	& 0.3	& 0.5 \\
		& Total			& 64.7  & 100  \\ 
\hline
$2^3S_1$	& $1^3S_1 +\pi\pi$	& $57\pm 7$ & 79.6 \\ 
		& $1^3P_2 + \gamma $ 	& 5.7	& 8.0 \\
		& $1P_1' +\gamma $ 	& 0.7	& 1.0 \\
		& $1P_1  +\gamma $ 	& 4.7	& 6.6 \\
		& $1^3P_0 + \gamma $ 	& 2.9	& 4.0 \\
		& $2^1S_0 + \gamma$ 	& 0.01	& $1\times 10^{-2}$ \\
		& $1^1S_0 + \gamma$ 	& 0.6	& 0.8 \\
		& Total			& 71.6	& 100 \\
\hline
\hline
$1^3D_3$	& $1^3S_1 +\pi\pi$	& 4.3  	& 5.2 \\ 
		& $1^3P_2 + \gamma $ 	& 78	& 94.8 \\
		& Total			& 82.3	& 100 \\ \hline
$1D_2'$		& $1^3S_1 +\pi\pi$	& 2.1 	& 2.5 \\ 
		& $1^1S_0 +\pi\pi$	& 2.2 	& 2.6 \\ 
		& $1^3P_2 + \gamma $ 	& 8.8	& 10.6 \\
		& $1P_1' +\gamma $ 	& 63	& 75.8 \\
		& $1P_1  +\gamma $ 	& 7	& 8.4 \\
		& Total			& 83.1	& 100 \\ \hline
$1D_2$		& $1^3S_1 +\pi\pi$	& 2.2 	& 2.4  \\ 
		& $1^1S_0 +\pi\pi$	& 2.1 	& 2.3 \\ 
		& $1^3P_2 + \gamma $ 	& 9.6	& 10.3 \\
		& $1P_1' +\gamma $ 	& 15	& 16.1 \\
		& $1P_1  +\gamma $ 	& 64	& 68.9 \\
		& Total			& 92.9	& 100 \\ \hline
$1^3D_1$	& $1^3S_1 +\pi\pi$	& 4.3 	& 4.6 \\ 
		& $1^3P_2 + \gamma $ 	& 1.8	& 1.9 \\
		& $1P_1' +\gamma $ 	& 4.4	& 4.7 \\
		& $1P_1  +\gamma $ 	& 28	& 29.9 \\
		& $1^3P_0 + \gamma $ 	& 55	& 58.8 \\
		& Total			& 93.5	& 100 \\
\hline \hline
\end{tabular}
\end{center}

\end{table}

\begin{table}
\caption{Partial widths and branching fractions
for strong and electromagnetic transitions (continued). 
\label{tab:br2}}
\begin{center}
\begin{tabular}{l l l r } \hline \hline
Initial \phantom{www}        & Final            	& Width  	& B.F. \\
 state          & state            	& (keV)  	& \ \ (\%)  \\ 
\hline
$2^3P_2$	& $1^3P_2 +\pi\pi$	& 1.0	& 1.3 \\
		& $2^3S_1 +\gamma$	& 55	& 70.3 \\
		& $1^3S_1 +\gamma$	& 14	& 17.9 \\
		& $1^3D_3 +\gamma$	& 6.8	& 8.7 \\
		& $1D_2' +\gamma$	& 0.7	& 0.9 \\
		& $1D_2 +\gamma$	& 0.6	& 0.8 \\
		& $1^3D_1 +\gamma$	& 0.1	& 0.1 \\
		& Total			& 78.2	& 100 \\ \hline
$2P_1'$		& $1P_1' +\pi\pi$	& 1.2	& 1.4 \\
		& $2^3S_1 +\gamma$	& 5.5	& 6.4 \\
		& $2^1S_0 +\gamma$	& 52	& 61.0 \\
		& $1^3S_1 +\gamma$	& 0.6	& 0.7  \\
		& $1^1S_0 +\gamma$	& 19	& 22.3 \\
		& $1D_2' +\gamma$	& 5.5	& 6.4 \\
		& $1D_2 +\gamma$	& 1.3	& 1.5 \\
		& $1^3D_1 +\gamma$	& 0.2	& 0.2 \\
		& Total			& 85.3	& 100 \\ \hline
$2P_1$		& $1P_1 +\pi\pi$	& 2.7	& 4.0 \\
		& $2^3S_1 +\gamma$	& 45	& 67.3 \\
		& $2^1S_0 +\gamma$	& 5.7	& 8.5 \\
		& $1^3S_1 +\gamma$	& 5.4	& 8.1 \\
		& $1^1S_0 +\gamma$	& 2.1	& 3.1 \\
		& $1D_2' +\gamma$	& 0.8	& 1.2 \\
		& $1D_2 +\gamma$	& 3.6	& 5.4 \\
		& $1^3D_1 +\gamma$	& 1.6	& 2.4 \\
		& Total			& 66.9	& 100 \\ \hline
$2^3P_0$	& $1^3P_0 +\pi\pi$	& 1.0	& 2.1 \\
		& $2^3S_1 +\gamma$	& 42	& 87.1 \\
		& $1^3S_1 +\gamma$	& 1.0	& 2.1 \\
		& $1^3D_1 +\gamma$	& 4.2	& 8.7 \\
		& Total			& 48.2	& 100 \\
\hline \hline
\end{tabular}
\end{center}

\end{table}

\begin{table}
\caption{Expected event rates for various decay chains at the Tevatron 
and the LHC.  The $B_c$ is assumed to decay to
$B_c\to J/\psi \pi \to 
\mu^+\mu^-\pi$ and $B_c\to J/\psi \ell \nu_\ell \to \mu^+\mu^-\ell 
\nu_\ell$ final states.  Details of these estimates
are described in the text.
\label{tab:events}}
\begin{center}
\begin{tabular}{l c c } \hline \hline
Decay Chain     & Tevatron            	& LHC  	 \\
\hline
$1^3S_1\stackrel{\gamma}{\to}B_c$  & $3.4\times 10^3$ & $3.4\times 10^5$ \\
$2^1S_0\stackrel{\pi\pi}{\to}B_c$  & $1.8\times 10^3$ & $1.8\times 10^5$ \\
$2^3S_1\stackrel{\pi\pi}{\to}1^3S_1\stackrel{\gamma}{\to}B_c$
				   & $1.6\times 10^3$ & $1.6\times 10^5$ \\
$2^3S_1\stackrel{\gamma}{\to}B_c$  & $16$ & $1.6\times 10^3$ \\
$1^3P_2\stackrel{\gamma}{\to}1^3S_1\stackrel{\gamma}{\to}B_c$  
				   & $6.7\times 10^2$ & $6.7\times 10^4$ \\
$1P_1'\stackrel{\gamma}{\to}B_c$  
				   & $5.9\times 10^2$ & $5.9\times 10^4$ \\
$1P_1\stackrel{\gamma}{\to}1^3S_1\stackrel{\gamma}{\to}B_c$  
				   & $5.5\times 10^2$ & $5.5\times 10^4$ \\
$1P_1\stackrel{\gamma}{\to}B_c$  
				   & $1.2\times 10^2$ & $1.2\times 10^4$ \\
$1^3P_0\stackrel{\gamma}{\to}1^3S_1\stackrel{\gamma}{\to}B_c$  
				   & $6.7\times 10^2$ & $6.7\times 10^4$ \\
$1^3D_3\stackrel{\pi\pi}{\to}1^3S_1\stackrel{\gamma}{\to}B_c$
				   & $3.5$ & $3.5\times 10^2$ \\
$1^3D_3\stackrel{\gamma}{\to}1^3P_2\stackrel{\gamma}{\to}1^3S_1
	\stackrel{\gamma}{\to}B_c$
				   & $64$ & $6.4\times 10^3$ \\
$1D_2'\stackrel{\pi\pi}{\to}1^1S_0$ & $1.7$ & $1.7\times 10^2$ \\
$1D_2'\stackrel{\gamma}{\to}1P_1'\stackrel{\gamma}{\to}B_c$ 
				& $45$ & $4.5\times 10^3$ \\
$1D_2\stackrel{\pi\pi}{\to}1^1S_0$ & $1.4$ & $1.4\times 10^2$ \\
$1D_2\stackrel{\gamma}{\to}1P_1'\stackrel{\gamma}{\to}B_c$ 
				& $9.5$ & $9.5\times 10^2$ \\
$1D_2\stackrel{\gamma}{\to}1P_1\stackrel{\gamma}{\to}B_c$ 
				& $8.2$ & $8.2\times 10^2$ \\
$1^3D_1\stackrel{\pi\pi}{\to}1^3S_1\stackrel{\gamma}{\to}B_c$
				   & $3.1$ & $3.1\times 10^2$ \\
$1^3D_1\stackrel{\gamma}{\to}1P_1'\stackrel{\gamma}{\to}B_c$
				   & $2.8$ & $2.8\times 10^2$ \\
$1^3D_1\stackrel{\gamma}{\to}1P_1\stackrel{\gamma}{\to}B_c$
				   & $3.6$ & $3.6\times 10^2$ \\
$2^3P_2\stackrel{\gamma}{\to}2^3S_1\stackrel{\gamma}{\to}B_c$  
				   & $2.3$ & $2.3\times 10^2$ \\
$2^3P_2\stackrel{\gamma}{\to}1^3S_1\stackrel{\gamma}{\to}B_c$  
				   & $72$ & $7.2\times 10^3$ \\
$2P_1'\stackrel{\gamma}{\to}B_c$  
				   & $90$ & $9.0\times 10^3$ \\
$2P_1\stackrel{\gamma}{\to}B_c$  
				   & $12$ & $1.2\times 10^3$ \\
$2P_1\stackrel{\gamma}{\to}2^3S_1\stackrel{\gamma}{\to}B_c$  
				   & $2.2$ & $2.2\times 10^2$ \\
$2^3P_0\stackrel{\gamma}{\to}2^3S_1\stackrel{\gamma}{\to}B_c$  
				   & $2.8$ & $2.8\times 10^2$ \\
\hline \hline
\end{tabular}
\end{center}

\end{table}

\section{Summary}

The primary purpose of this paper is to calculate $B_c$ masses and 
radiative transitions 
in the relativized quark model.  For the most part the 
mass predictions are consistent with other models,
within the accuracy of these models.
The largest discrepancy in predictions is for the 
triplet-singlet mixing angles.  
This has implications for transitions between 
states so can be tested with appropriate measurements.  Combining the 
BR's we have calculated with $B_c$ production cross sections from the 
literature we see that the $1S$, $1P$, and $2S$ states should be 
produced in sufficient numbers to be observed at the Tevatron.  
With the higher statistics of the LHC, it should also be possible 
to observe the $1D$ and $2P$ states.  It will be a significant 
experimental challenge to extract the signals for these states from 
the large background but their observation 
would add considerably to 
our knowledge of quarkonium spectroscopy and discriminate between the 
various models that exist in the literature.

\acknowledgments

The author thanks Vaia Papadimitriou and Manuella Vincter
for helpful communications.
This research was supported in part 
the Natural Sciences and Engineering Research Council of Canada.



\begin{thebibliography}{99}
\bibitem{salpeter}
The relativistic 
kinetic energy is actually that of the spinless Salpeter equation.
\end{thebibliography}
\end{document}